\documentclass[article, nofootinbib]{revtex4-2}
\usepackage{multirow, amsmath, hyperref, graphicx, booktabs}
\usepackage{dcolumn}
\usepackage[dvipsnames]{xcolor}
\usepackage[final]{microtype}
\usepackage{amsmath}
\usepackage{amssymb}
\usepackage{subcaption}
\usepackage[normalem]{ulem}
\hypersetup{
colorlinks=true,
citecolor=cyan,
linkcolor=magenta,
urlcolor=NavyBlue,
allbordercolors={0 0 0},
pdfborderstyle={/S/U/W 1}
}
\usepackage{orcidlink}
\newcommand{\IWF}{Space Research Institute, Austrian Academy of Sciences, Schmiedlstrasse 6, 8042 Graz, Austria}
\begin{document}

\title{Machine learning-based classification \\for Single Photon Space Debris Light Curves}

\author{\sc{Nadine M. Trummer}}
\email{nadinemaria.trummer@oeaw.ac.at}
\affiliation{\IWF}
\author{\sc{Amit Reza}}
\email{amit.reza@oeaw.ac.at} \affiliation{\IWF}
\author{\sc{Michael A. Steindorfer}}
\email{michael.steindorfer@oeaw.ac.at}
\affiliation{\IWF}
\author{\sc{Christiane Helling}}
\email{christiane.helling@oeaw.ac.at} \affiliation{\IWF}
\begin{abstract}
\textbf{Abstract} 
The growing number of man-made debris in Earth’s orbit poses a threat to active satellite missions due to the risk of collision. Characterizing unknown debris is, therefore, of high interest. Light Curves (LCs) are temporal variations of object brightness and have been shown to contain information such as shape, attitude, and rotational state. Since 2015, the Satellite Laser Ranging (SLR) group of Space Research Institute (IWF) Graz has been building a space debris LC catalogue. The LCs are captured on a Single Photon basis, which sets them apart from CCD-based measurements. In recent years, Machine Learning (ML) models have emerged as a viable technique for analyzing LCs. This work aims to classify Single Photon Space Debris using the ML framework. We have explored LC classification using k-Nearest Neighbour (k-NN), Random Forest (RDF), XGBoost (XGB), and Convolutional Neural Network (CNN) classifiers in order to assess the difference in performance between traditional and deep models. Instead of performing classification on the direct LCs data, we extracted features from the data first using an automated pipeline. We apply our models on three tasks, which are classifying individual objects, objects grouped into families according to origin (e.g., GLONASS satellites), and grouping into general types (e.g., rocket bodies). We successfully classified Space Debris LCs captured on Single Photon basis, obtaining accuracies as high as 90.7\%. Further, our experiments show that the classifiers provide better classification accuracy with automated extracted features than other methods. \\ 

\end{abstract}

\maketitle 


\section{Introduction}
\label{sec:intro}
Since the launch of the world's first satellite in 1957, the population of man-made objects in Earth's orbit has experienced a continuous growth \citep{ESAReport2023}. 
Today, non-functional debris make up the majority of in-orbit objects\citep{ESASpaceDebris}. 
These so-called space debris traverse space with high relative velocities with respect to other objects in orbit (7-8 km/s), which renders them a threat to active satellite missions due to risk of collision \citep {Schildknecht2007}. Although each potential collision event poses an issue on its own, the problem goes far beyond the damage to individual satellites.
In their work, \cite{kessler1978} predict what became known as the Kessler-Syndrome: fragmentation caused by collisions increases the risk of future collisions, eventually resulting in a dense debris belt around Earth that would cause significant problems for future space missions. 
It is of high interest to characterize shape, surface properties, attitude state, and spin of debris objects since these directly affect orbit propagation \citep{Fan2020}.
Inferring these properties, however, is no trivial task. Due to their small size and atmospheric attenuation, not all space objects are optically resolvable, making characterization challenging.

Light Curves (LCs) are temporal variations of an object's observed brightness and can be measured even if the object itself is not resolvable. In the context of space debris, LCs have been shown to contain information about object properties such as shape \citep{Silha2021}, attitude \citep{Kucharski2021} and rotational state \citep{Silha2018}. 
LCs of space debris are usually obtained by extracting pixel brightness from a sequence of non-resolved CCD-based telescope images  \citep{Silha2018, allworth2021}.
The data in this work, however, has been gathered differently. We utilize Single Photon LCs, which are collected by directly counting the incoming photons using a Single Photon Avalanche Diode (SPAD) detector, thereby bypassing the step of brightness extraction. 
To infer object properties from LCs, inversion techniques can be applied (e.g., \citep{burton2021}). These techniques, albeit promising, require the estimation of a large number of parameters and may not be practicably applicable to extensive data catalogues \citep{Linares2020}.
In recent years, machine learning (ML) has emerged as a viable alternative technique for LC analysis. Recent efforts have focused on classifying space debris by applying neural networks (NN) \citep{wu2018development}, e.g., Convoluational Neural Network (CNN) \citep{albawi2017understanding, wu2017introduction, gu2018recent}, using LCs as input for training \citep{allworth2021, Furfaro2016, Furfaro2019, Linares2020}.
The strength of NN lies in their capability of learning complex, non-linear decision boundaries that allow them to outperform traditional ML algorithms (e.g., Support Vector Machine (SVM) \citep{hearst1998support} and Decision Trees \citep{quinlan1987simplifying}). CNN, in particular, show great potential in solving classification problems as they can extract local and global features during training. NN-architectures, however, may comprise millions of parameters. Hence, a large volume of training data is required for them to unfold their full potential \citep{LeGuennec2016, Wen2020, Iwana2020}. 
Albeit their promising performance, applying CNN to space debris classification suffers from the limited quantity of well-labeled, high-quality data available. Past studies made efforts to address the issue by introducing simulated data \citep{allworth2021,Linares2020} as part of training.

The presented work aims to explore the suitability of Single Photon Space Debris Light Curves for space debris classification using the ML framework. The remainder of the paper is organized as follows: Section \ref{sec:approach} describes our study approach by introducing the \textbf{IWF} \textbf{SPARC} (\textbf{S}ingle \textbf{P}hoton Light Curve \textbf{A}nd Laser \textbf{R}anging \textbf{C}atalogue) catalogue as well as the ML framework and the models employed in this study. Section \ref{sec:data} gives details on data acquisition, pre-processing and labeling of LCs, which is required for the training process. Section \ref{sec:methods} presents the methods used in this work. We report our results in the follow-up Section \ref{sec:results}. In Section \ref{sec:conclusion}, we summarize our findings and discuss opportunities as well as limitations of the presented classification scheme. Further, we refer to some ideas for expanding on our current framework studies.

\section{Approach}
\label{sec:approach}
The objective of this work is to classify space debris utilizing data from IWF SPARC, thereby demonstrating Single Photon Space Debris LCs are a viable alternative to non-resolved CCD images. This section will outline our study's approach. After introducing IWF SPARC, we will describe the ML framework and will give an overview of the strategies used to classify the LCs.

\subsection{Study Dataset}
This investigation  utilizes data from IWF SPARC, which is a collection of LC measurements conducted by IWF Graz over the time-span from 2015 to February 2023. It includes more than 6500 LCs covering over 700 individual satellites and space debris objects. Between one and 243 LC measurements are available per object, with around 60\% of these objects being represented by three or less LCs. The space objects presented in IWF SPARC can be broadly categorized into active satellites (e.g., Sentinel-3a or geodetic satellites \citep{pearlman2019laser} such as LAGEOS-1), defunct or decommissioned satellites (e.g., ENVISAT or TOPEX/Poseidon), and debris (e.g., rocket bodies left behind during launch processes). Each object has a NORAD (North American Aerospace Defense) Catalog Number, which is an internationally recognized identifier that allows to retrieve additional object information from public sources (e.g., current and historic orbit data\footnote{https://www.space-track.org}).
\begin{figure}
\centering
\begin{subfigure}[b]{\textwidth}
\centering
\includegraphics[width=\textwidth]{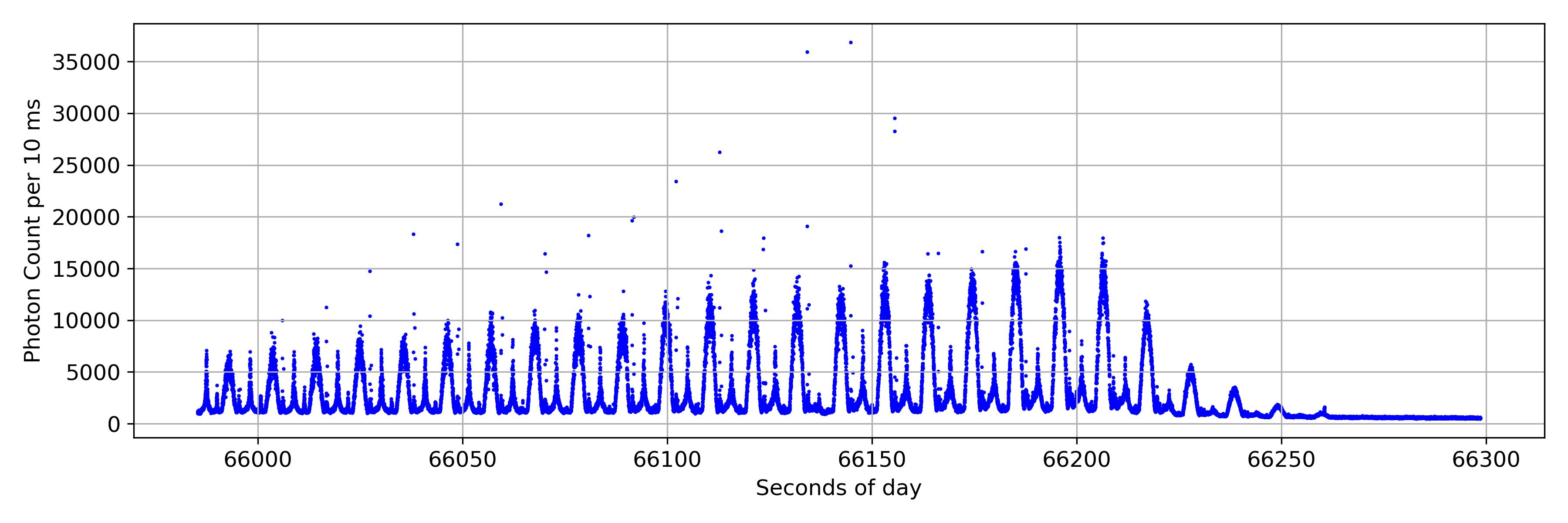}
\caption{A LC captured on 2017/1/27. Up until the 66200 s mark, there is an overall increase in amplitude to be observed, which is caused by changing observation geometry. Further, the LC contains a repeating pattern of a main peak followed by a secondary peak of lower amplitude. This pattern likely corresponds to individual surfaces of the target. After the 66200 s mark, the signal strength decreases rapidly, which is due to the target exiting the telescope's field-of-view or vanishing in Earth's shadow.}
\label{fig:topex_a}
\end{subfigure}
\hfill
\begin{subfigure}[b]{\textwidth}
\centering
\includegraphics[width=\textwidth]{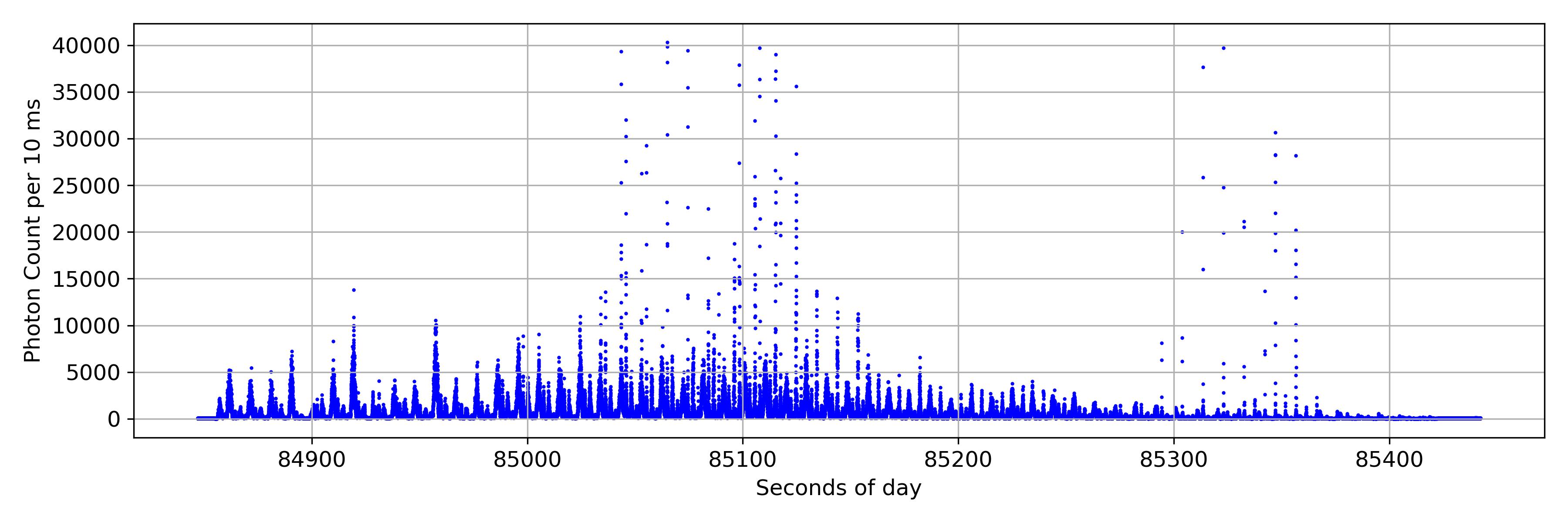}
\caption{A LC captured on 2022/7/8. In addition to a regular pattern comparable to the one depicted in \ref{fig:topex_a}, glints are visible from 85050 s to 85150 s as well as from the 85300 s mark onward. Glints are narrow spikes in brightness. They only comprise a couple of data points, but exhibit high amplitude compared to the regular signal (here, as high as 40000 counts per 10 ms).}
\label{fig:topex_b}
\end{subfigure}
\caption{Raw LC measurements of TOPEX/Poseidon (NORAD 22076), a decommissioned satellite in LEO with a spin period of 10.73 sec as per 2016 \citep{kucharski2017}.}
\label{fig:topex}
\end{figure}

Figure \ref{fig:topex} depicts two raw LC measurements of TOPEX/Poseidon (NORAD 22076), a decommissioned satellite in Low Earth Orbit (LEO) with a known spin period of 10.73 s as per 2016 \citep{kucharski2017}. The measurement in Figure \ref{fig:topex_a} shows a repeating pattern of a peak followed by a secondary peak with lower amplitude. The pattern is a mixture of specular and diffuse reflections off the target \citep{Kucharski2021}. The photon count (i.e., signal amplitude) increases in proportion to the fraction of illuminated surface area visible to the observer. In case of a rotating object, this fraction will change over time.
With a spin period of around 10 s, the decommissioned satellite TOPEX/Poseidon is one of the fastest rotating objects in IWF SPARC. Most objects rotate significantly slower (e.g., ENVISAT with 134 s \citep{kucharski2014}); \citep{kirchner2017determination} analysed 44 defunct GLONASS satellites and found rotation periods ranging from 8 s to over 400 s, with 41 of these objects being represented in IWF SPARC. This information helps provide a baseline estimate for how long LC measurements need to be in order to provide meaningful information on a target, since ideally, at least one full rotation is captured. The mean duration of our LC measurements is 269 s, while the median duration is 187 s. 

\begin{figure}
\centering
\includegraphics[width=\textwidth]{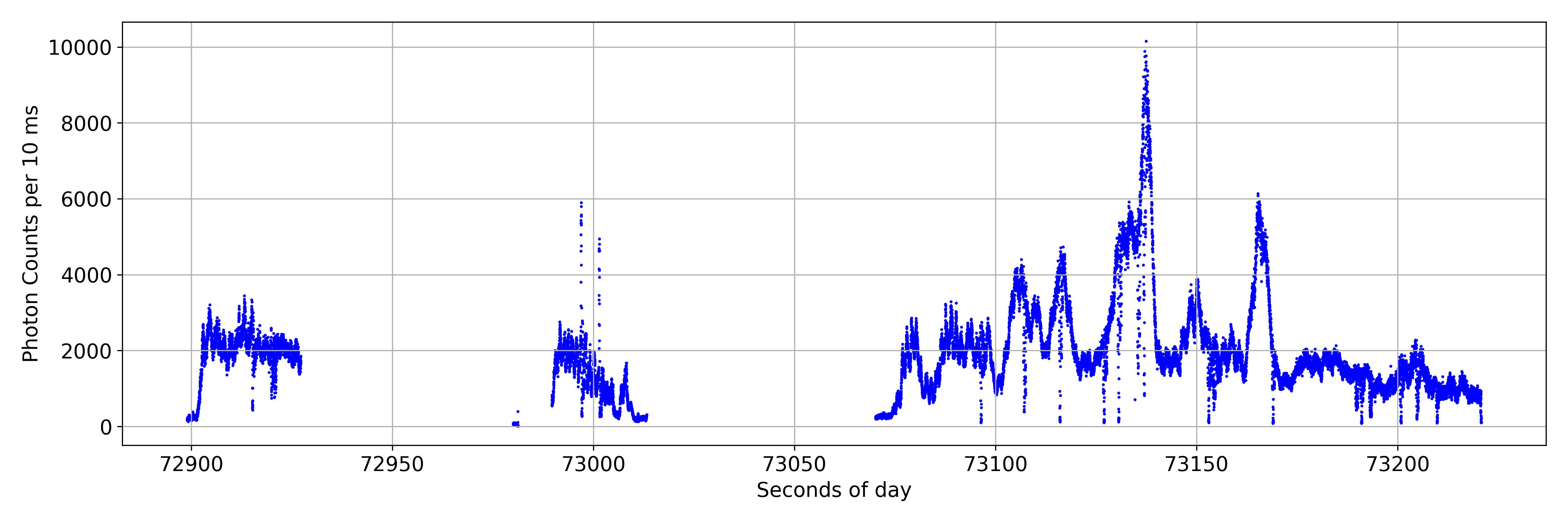}
\caption{A LC of satellite Sentinel-3b (NORAD 43437), captured on 2022/4/30. The first half of the measurement experiences interruptions in the time domain. Though the measurement file can be considered faulty, the latter half of the measurement can still potentially be used for analysis.}
\label{fig:faulty}
\end{figure}

The overall quality of the dataset is inconsistent. Sudden jumps in brightness (caused by e.g., a background star crossing the field-of-view), temporal gaps within the data (see Figure \ref{fig:faulty}) or the total absence of a signal (noise-only) are examples for undesired behavior. However, LCs also contain signatures that could be misinterpreted as noise or outliers, as the following example will show. Figure \ref{fig:topex_b} shows a LC of TOPEX/Poseidon containing what we call glints. These are narrow spikes in photon count that only comprise a couple of data points but are high in amplitude compared to the rest of the signal. We interpret these glints as moments of specular reflection. This type of reflection occurs on smooth, shiny surfaces (e.g., solar panels, antennas), for the case that the angle of incidence equals the angle of reflection with regards to the surface normal (Snell's law), resulting in mirror-like reflection of sunlight on the target's surface. If the observer is looking at the surface at the aforementioned angle, they perceive this specular reflection of sunlight as a sharp increase in object brightness, which is represented in the measurement by a steep, narrow spike in photon count. Glints as depicted in Figure \ref{fig:topex_b} should therefore not be mistaken to be outliers or noise.

Though a major strength of ML is its ability to work with incomplete datasets, it was decided to select a well-behaved subset of our data for this study, since this work is interested in investigating how ML approaches perform on our Single Photon LCs. The subset was chosen through visual inspection and comprises LCs that display a clear signal (i.e., object is in field-of-view and rotating), have at least a length of 100 s and do not exhibit any of the aforementioned quality faults.  
An exception to this rule are geodetic satellites, which are expected to display smoother curves due to their spherical geometry. The geodetic satellite Ajisai is special in this regard, since its surface is covered in mirrors instead of retro-reflectors and therefore continuously shows glints.  To give an example, both curves depicted in Figure \ref{fig:topex} are suitable. To avoid discarding data, we extracted portions of LCs, where possible (e.g., the last third of Figure \ref{fig:faulty}). The resulting data subset applied in this study consists of $852$ LCs representing $150$ individual objects.

\subsubsection{Machine Learning Framework}

This study utilizes multiple ML models for analyzing LCs. The ML standard procedure necessitates building, training, testing, and deploying ML models to classify light curve data. Setting up the training, validation, and testing process is crucial for developing robust models. The training set is a subset of the available data used to train ML models. Training a ML model involves optimizing the model parameters to minimize a chosen loss function. This also includes the optimization of hyperparameters, which are ML model parameters that cannot be changed during training, such as the number of layers in a NN \citep{yu2020hyper}. The validation set evaluates the model's performance during training and is used for tuning hyperparameters. The testing set assesses the final performance of the trained model after tuning and model selection. For the training of ML models, the available dataset is typically divided into training, validation, and testing sets using a predefined ratio  for splitting the dataset (e.g., 80\% of data being used for training, 10\% for validation, and 10\% for testing). 

\begin{figure}[ht!]
\centering
\includegraphics[width=0.5\textwidth]{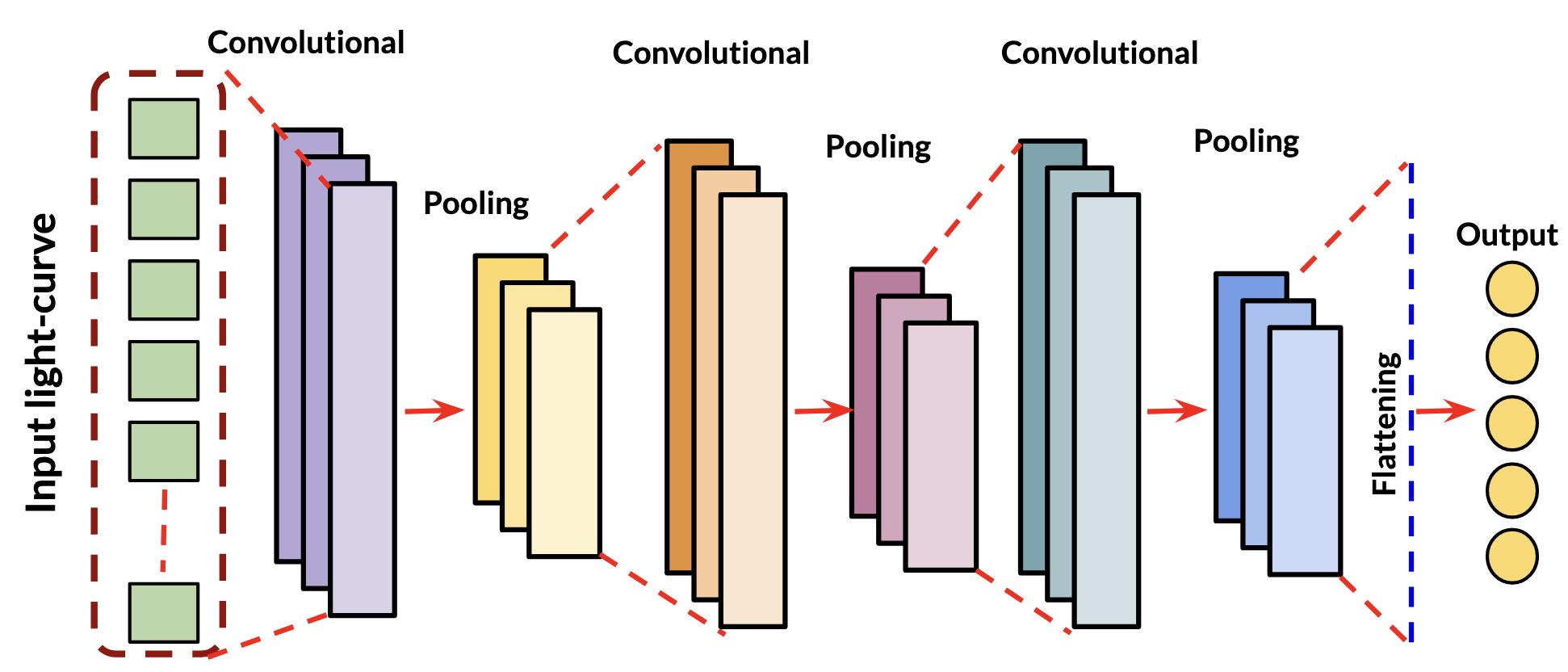}
\caption{The schematic diagram of the one-dimensional CNN architecture used for the LCs classification problem. Three convolution and max pooling layers are used between the input and output layers. All the layers used a fixed kernel and filter size of $64$ and $3$, respectively. The tanh activation function is used for convolution layers, whereas the softmax activation function has been chosen for output layers. Batch normalization is used between the convolution and max pooling layers. }
\label{fig:cnn_diag}
\end{figure}

NN models have yielded promising results in the space debris light curve classification task \citep{Allworth2020}. Currently, we have access to a limited data set size, so we have decided to investigate the performance of other algorithms besides NN models. This study employs a total of four different ML models. They are k-Nearest-Neighbour classifier ($\texttt{k-NN}$) \citep{cunningham2021k}, Random Forest Classifier (RDF) \citep{pal2005random}, XGBoost (XGB) \citep{chen2016xgboost} and Convolutional Neural Network (CNN) \citep{albawi2017understanding}: 
\begin{itemize}
\item \textbf{\texttt{k-NN}} is an ML classifier that provides top-$\texttt{k}$ nearest neighbours to a given query point by measuring the distance between the points. The label of the query point is then predicted by the average of the closest neighbours. In this study, \texttt{k-NN} needs to perform with light curve time series data, which could be mathematically conceptualized as higher dimensional data points. Hence, we can use Euclidean Distance or Dynamic Time Warping (DTW) as a distance measure to identify the nearest neighbours.

\item The \textbf{RDF} classifier creates multiple decision trees from the random subsets of the data and features. Each decision tree learns the features of the data and classifies the data accordingly. Due to the random choice of the data and features, each tree interacts with a different number of features and samples of the data. Thus, the prediction of each tree could vary. The final prediction result is obtained based on the majority voting scheme. 
\item \textbf{XGB} is a decision tree-based ensemble learning method that builds a robust predictive model by sequentially adding weak learners. The minimization of a loss function is performed at every stage by optimizing the predictions. 
\item \textbf{CNN} is a deep learning model designed explicitly for classifying structured data (e.g., images). The introduction of CNN revolutionized the field of computer vision and is widely used for various tasks, including classification, object detection, and image segmentation. The CNN architecture mainly contains Convolutional, Pooling, and Fully connected Layers. 
\begin{enumerate}
\item Convolutional Layers: CNN architecture consists of multiple convolutional layers that learn spatial hierarchies of features from input images or time series. Each convolutional layer applies filters (kernels) to the input, capturing local patterns such as edges, textures, and shapes. Thus, convolutional layers help extract hidden local features from the input.  
\item Pooling Layers: Pooling layers (e.g., max pooling) are generally inserted after convolutional layers to reduce spatial dimensions and extract dominant features while preserving spatial invariance.
\item Activation Functions: Activation functions (e.g., ReLU, ELU, Tanh) are non-linear functions applied after convolutional and pooling layers to introduce non-linearity. It enables the network to learn complex relationships in the data.
\item Fully Connected Layers: After several convolutional and pooling layers, CNN ends with fully connected layers followed by a softmax activation function for the classification problem. These layers integrate the learned features and make final predictions.
\end{enumerate}
The training in the CNN model depends mainly on the forward, backward propagation processes and regularization. The forward propagation connects the input and output layers with the training data. That means the input data has been passed through the design CNN architecture, and in the output layers, it predicts the corresponding classification labels. If the predicted classification labels are not the same as the known labels, then it is required to perform backpropagation. This is known as loss calculation, and cross-entropy loss is commonly used for classification cases. Mathematically, the cross entropy loss is defined as
\begin{equation}
\mathcal{L}(\theta) = \frac{1}{N}{\sum_{i = 1}^{N} {- \log p(\tilde{y} = \tilde{y}_{i} | x_{i}, \theta)} }
\end{equation}
The backpropagation process involves several operations: gradient calculation, parameter update, backward pass through the layers, update weights of the fully connected layers, and update of the convolutional filters. The CNN architecture used for this work is shown in Figure \ref{fig:cnn_diag}. 
\end{itemize}

\textbf{Strategies for light-curve classification:} 
NN models are designed to operate and accurately classify complex time series data. However, other ML algorithms (e.g., Decision Trees, \texttt{k-NN}) can grapple to obtain similar accuracy while training time series data with complex features. Their performance could be enhanced by passing featured-engineered data as input instead of directly trained on raw time series data. In the domain of time series classification, distance-based and feature-based representations of data are well-established methods \citep{Abanda2019} to improve the accuracy of the classifiers. We have also integrated these methods with all four ML models. We have performed feature engineering to extract the essential and relevant features and then use them with RDF, XGB, and CNN classifiers. We have used ED and DTW (distance)-based on a \texttt{k-NN} classifier. Details of these approaches are presented in Section \ref{sec: methods}.

\begin{figure}[ht!]
\centering
\includegraphics[width=0.5\textwidth]{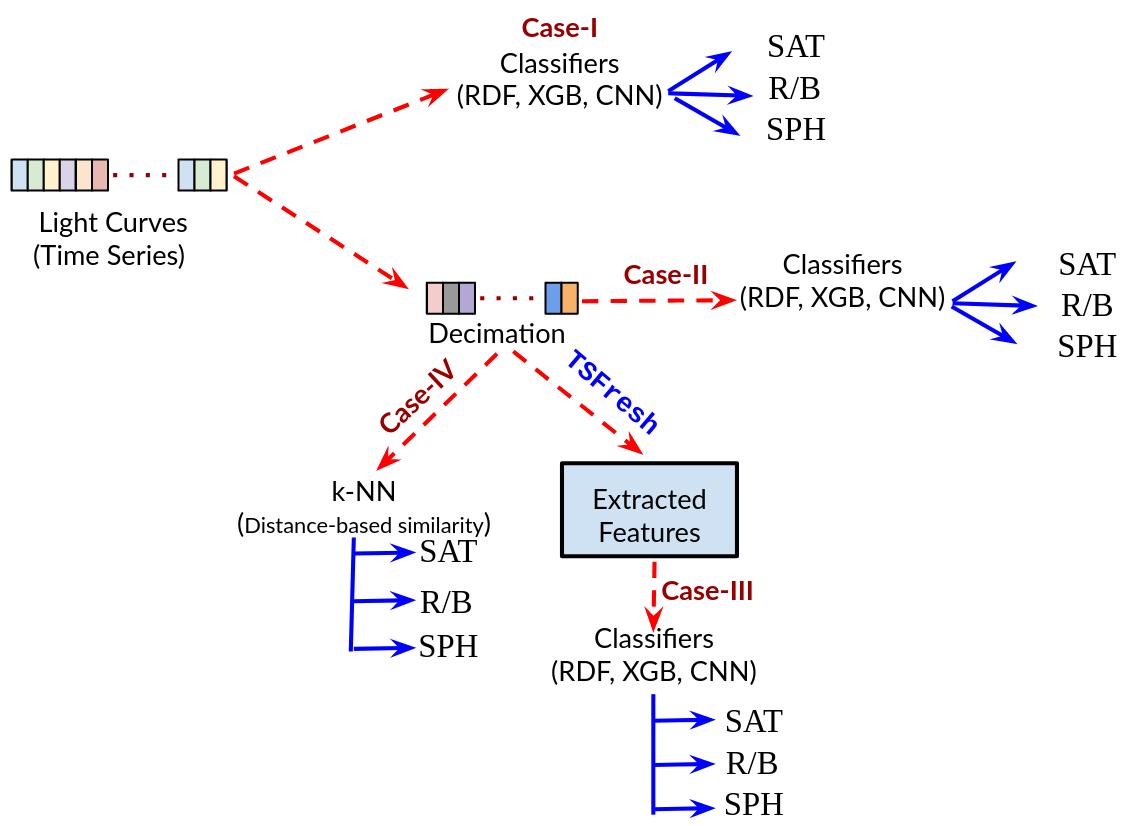}
\caption{The schematic representation of methods used in this work at example of the \textit{Types} subset.}
\label{fig:Methods_workflow}
\end{figure}

The possibility of training raw and featured engineered data with ML models allows us to examine several model configurations. Figure \ref{fig:Methods_workflow} summarizes all possible strategies used in this study. 
Case-I describes the training of ML classifiers on data with original resolution (raw data). Case-II involves decimation (downsampling) of LCs and then learning the decimated LCs to train the classifiers. Case-III extracts the automated features, and the extracted features are then used to train classifiers. Case-IV uses distance-based similarity measurement to train the \texttt{k-NN} classifier using the downsampled LCs. 
At this point, it is essential to mention that the extracted feature and distance-based methods could also work with LC time series with original resolution. However, the computational cost of extracting features or computing similarity using distance measures would be high. Therefore, we have incorporated extracted feature/ distance schemes with a downsampled version of LCs. We have investigated the impact of downsampling in detail in the result section (see Section \ref{sec:preprocess}) before deciding on the downsampling rate. Figure \ref{fig:Methods_workflow}  demonstrated the classification on a specific data subset called \textit{Types} (which includes three classes SAT, R/B, SPH) to demonstrate our approach. We have tested these strategies for our other classification tasks (i.e., \textit{Uniques}, \textit{Families}) as well. Each task draws from the aforementioned data subset of $852$ LCs but applies different labeling schemes. We give details on the classification tasks considered in this study in Section \ref{sec:onthology}.

\textbf{Evaluation procedure for ML models:} We evaluated the performance of our models using k-fold cross-validation, which provides more robust accuracy by repeatedly splitting the data into training and validation sets. The data is divided into k subsets (folds), and the model is trained k times, each time using a different fold as the validation set and the remaining folds as the training set. In addition to classification accuracy, we provide standard ML metrics (e.g., precision, recall, f1-score, specificity, and false positive rate) \citep{erickson2021magician, grandini2020metrics} to evaluate the performance of models.

\textbf{Confusion Matrix:} Similar to hypothesis testing, a confusion matrix (CM) representation helps to understand the probabilities of different prediction outcomes of the trained model. For example, in binary classification problems, CM represents predicted against true classes and allows the identification of the frequency of true positives ( correct class), true negatives (correct class), false positives (type-I error), and false negatives (type-II error) in a prediction. The counts of these variables are used to calculate several metrics (e.g., True positive rate, True negative rate, False positive rate) to evaluate the performance of a specific predictive ML model. 
For multi-class classification with $p$ classes, the rows and columns of the confusion matrix $C_{p \times p}$ represent the actual and predicted classes, respectively. Each cell ($C_{ij}$) of the matrix represents the number of instances of class-$i$ that were predicted as class-$j$. Each cell ($C_{ij}$) of the matrix represents the number of instances of class-$i$ that were predicted as class-$j$. In a well-performing model, the diagonal cells (representing correct predictions) should have higher values than off-diagonal elements. High values of off-diagonal cells indicate a high rate of misclassifications, and examining these can provide insight into specific classes that are being confused with others. \\

\textbf{Key metrics:} From the confusion matrix, various vital metrics can be derived that are indicators of the model's performance. We highlight the metrics that were used in our study. 
\begin{itemize}
\item \textbf{Precision:} Precision measures the accuracy of the positive predictions, meaning that out of all predicted instances as positive, how many are truly positive. Precision can be defined as the function of True positives (TP) and False positives (FP) for binary classification, i.e., 
\begin{equation*}
\text{Precision} = \frac{\text{TP}}{\text{TP} + \text{FP}}
\end{equation*}
Extending it for multi-class classification, the precision for a specific class-$i$ is defined as $\frac{C_{ii}}{\sum_{j = 1}^{p}{C_{ji}}}$
\item \textbf{Recall:}
Recall measures the possibility of finding all the positive instances. Mathematically, it can be considered as a function of TP and false negatives (FN) as follows: 
\begin{equation*}
\text{Recall} = \frac{\text{TP}}{\text{TP} + \text{FN}}
\end{equation*}
\item \textbf{Specificity:}
Specificity, the opposite of recall, measures the proportion of actual negative instances correctly identified by the model.
\begin{equation*}
\text{Specificity} = \frac{\text{TN}}{\text{TN} + \text{FP}}
\end{equation*}
\item \textbf{$F1$ score:}
The $F1$ score score is calculated as the harmonic mean of precision and recall. The value of the F1 score lies between 0 and 1, with 0 indicating poor performance and a value of 1 representing perfect performance.
\begin{equation*}
\text{F}_{1} = 2 \times \frac{\text{Precision} \times \text{Recall}}{\text{Precision} + \text{Recall}} = \frac{\text{TP}}{\text{TP} + \frac{1}{2} (\text{FN} + \text{FP})}
\end{equation*}

\item \textbf{False Positive Rate:} The false positive rate (FPR) measures the probability of producing a false alarm, indicating a positive result when the actual value is negative.
\begin{equation*}
\text{FPR} = \frac{\text{FP}}{\text{FP} + \text{TN}} = 1 - \text{Specificity}
\end{equation*}
\end{itemize}

\section{Data}
\label{sec:data}
This section is concerned with describing the technical background of the observational method used to obtain the Single Photon Space Debris Light Curves in IWF SPARC.  We further give details on data pre-processing and lastly, explain how the labels of our data have been obtained.
\subsection{Acquisition}
\label{sec:acquisition}
IWF Graz routinely performs Satellite Laser Ranging (SLR) \citep{Wilkinson2019} to cooperative targets as well as space debirs \cite{Steindorfer2020} at Lustbühel Observatory Graz, Austria. SLR is a technique used to precisely measure the distance between a ground station and a space object. The object is targeted with monochromatic light ($\lambda_{\mathrm{\mathrm{Laser}}} = 532$ nm), provided by a pulsed laser source. The signal reflected from the target is gathered by a receiving telescope and routed towards a detection package, where  $\lambda_{\mathrm{Laser}}$ is separated from the rest of the spectrum using a dichroic mirror. By measuring the photon time-of-flight, the distance between target and observer can be determined. For details on the SLR routine see \cite{Steindorfer2020}.
While $\lambda_{\mathrm{Laser}}$ is used for the purpose of SLR, the residual spectrum can be used to obtain a LC of the target if at nighttime the object is illuminated by the sun (i.e., not obscured by Earth's shadow). In our setup, the residual spectrum of $\lambda = 750$ to $900$ nm is routed to a Single Photon Avalanche Diode ($\tau$-SPAD). The number of detected photons is a measure for object brightness, or in other words, how much sunlight is reflected from the object's surface towards the observer. The photons are counted through a Field Programmable Gate Array (FPGA) \citep{trimberger2012field} board and sorted into time bins of $\Delta t_{bin}=10$ ms, resulting in a constant sampling frequency of $f_n=$ 100 Hz \citep{Steindorfer2015}. The internal time frame of the FPGA is synchronized to Universal Coordinated Time (UTC) via GPS, therefore each bin is also time tagged. Measurements are stored in files stating the measurement epoch in seconds of day and the corresponding photon count. The high sampling frequency grants the system the capability to capture moments of specular reflection (glints, as described in figure \ref{fig:topex_b}) while the generated data quantities remain small (a few thousand kilobytes per file) compared to camera images, since only one pixel is involved in the measurement process. 
\subsection{Pre-Processing}\label{sec:preprocess}
The subset of 872 LCs chosen for this study has undergone visual quality inspection, however, there is still one major problem that needs addressing. Individual LC measurements have different lengths. Generally, classification models require the input to be of uniform shape (i.e., each LC input needs to be passed to the model as an array of exactly the same, pre-defined dimension). For this reason, our chosen 852 LCs were further sliced into segments of 100 s. This method, also referred to as window slicing, can be considered a form of data augmentation \citep{LeGuennec2016}. The chosen length of 100 s is the result of a trade-off. Shorter windows (e.g., 30 s) would allow for more segments to be generated, however, the segment length should be long enough to represent at least one full rotation of most objects in order to capture enough information. 
Apart from boosting available training samples, segmentation also is in this case also valid from a physical point of view. Different slices along a pass trajectory represent different lighting conditions (as illustrated by e.g., the occasional appearance of glint-patterns in figure \ref{fig:topex_b}). During a pass, the phase angle bisector (vector between observer-target-Sun) will change due to orbit geometry, which in return affects the illumination of the object, an effect that is most severe in LEO.  
The chosen length of 100 s aims to exploit the data augmentation aspect of the slicing while still ensuring LCs remain meaningful. This results in 852 being split into a total of 1141 LCs. 

\textbf{Discussion of Noise:} In this study, we did not take any additional steps to pre-process the data with regards to noise, since the chosen subset of LCs mitigates the issue by only conisdering LCs that display clear signal in relation to this background noise. We expect noise from various sources to be present in our data, namely the detector specific dark count rate ($<$100 counts per second for our device), the signal from the background sky as well as light pollution from the nearby city of Graz. Quantifying such noise is difficult, because conditions vary from night to night and are also depended on the pointing direction of the telescope, which changes while tracking the target across the sky. Further, applying any kind of noise-related pre-processing to the data needs to be done so with caution, since signatures like glints (as described in Figure \ref{fig:topex_b}) should not be repressed. Future analysis will require a more thorough discussion of noise, since we want to be able to feed the algorithm with less than ideal data. Lastly, backscattered light from the SLR laser is not an issue here, since the wavelength of 532 nm is not considered in LC measurements.

\subsection{Light Curve Categories}\label{sec:onthology}
Determining ground truth labels for LCs remains one of the fundamental challenges in this research field \citep{allworth2021}, since it requires prior knowledge of the object. Different studies investigate different classification tasks (such as differentiating rocket bodies or observed objects \citep{allworth2021}, differentiating types such as satellites, debris and rocket bodies \citep{Linares2020}).
The choice of labels is governed by the limitations and opportunities a data catalogue provides (e.g., a catalogue heavy on rocket bodies might be used to differentiate between different types of rocket bodies).
In this study, we apply three labeling schemes called \textit{Uniques}, \textit{Families} and \textit{Types}. All these are data subsets that draw from IWF SPARC. We do this in an attempt to provide comparability with the various labeling schemes introduced in other studies while also analysing the LCs of our IWF SPARC catalogue with various degrees of abstraction.
Tables \ref{tab:onthology_uniques}, \ref{tab:onthology_families} and \ref{tab:onthology_types} summarize the schemes and give an overview of the included objects as well as the number of LCs they contribute. The application of ML models to classify one of these data subsets will be referred to as a task (e.g., the \textit{Uniques}-task is concerned with classifying the \textit{Uniques} subset). In the following, we will discuss each labeling scheme in detail and give reasoning for our choices. 

\textbf{Uniques:} The most basic way of labeling a LC is to assign it to the individual object it represents. For this task, 8 out of 150 individual objects have been selected based on the overall number of LCs available for each of them in order to ensure a large enough ensemble to allow classification. Out of these 8 objects, Jason-2 and TOPEX/Poseidon contribute the most LCs to this subset, with 214 LCs and 132 LCs respectively, followed by the geodetic satellite Ajisai with 94 LCs. A summary of this subset is given in Table \ref{tab:onthology_uniques}. 

 Lares and Ajisai are examples for geodetic satellites, exhibiting a spherical shape. Jason-2, TOPEX/Poseidon as well as Glonass92 and Glonass67 are typical examples for dysfunctional boxwing satellites. Envisat is another defunct satellite but does differ in size and solar panel arrangement from the aforementioned satellites. Lastly, ATLAS 5 is a rocket body and is assumed to be of cylindrical shape with a nose tip and a cone. Ajisai, Lares, Envisat, Jason-2 and TOPEX/Poseidon are objects of the Low Earth Orbit (LEO, orbit height $\leq$ 2000 km). Glonass67 and Glonass92, both defunct satellites formerly being part of a GNSS (Global Navigation Satellite System), reside at  approximately 19.000 km (Medium Earth Orbit, MEO). The ATLAS 5 rocket body is unique since it is on a highly elliptical orbit around Earth (perigee 3.220 km, apogee 35.242 km), orbit heights at time of measurement therefore vary.

Challenges of this labelling scheme are the severe class imbalances as well as the overall low number of samples per class. However, an advantage of this scheme is that obtaining labels for the data is straight forward and unambiguous. Training a classifier on this problem has limited real-world applicability. It could, however, be employed to assess the similarity of unknown space debris with any of the catalogued objects, though the nature of that similarity would remain object of future study.
\begin{table}[ht!]
\caption{\textit{Uniques} subset (8 classes) with 675 LCs total.}
\centering
\begin{tabular}{|c|c|c|c|}
\hline
\textbf{Description} & \textbf{Label} &\textbf{LCs} & \textbf{\% of subset}\\
\hline
Jason-2 & JA2 & 214 & 31.7\\
TOPEX/Poseidon & TOP & 132 & 19.5 \\
AJISAI & AJI & 94 & 13.9\\
ENVISAT & ENV & 70 & 10.4\\
GLONASS092 & G92 & 51 & 7.6\\
GLONASS067 & G67 & 49 & 7.3\\
LARES & LAS & 36& 5.3\\
ATLAS 5 R/B & A05 & 29 & 4.3\\
\hline
\end{tabular}
\label{tab:onthology_uniques}
\end{table}

\textbf{Families:} Second, we sorted objects, were applicable, into object families. For example, we grouped all objects belonging to the GLONASS constellation into a group, as well as spherical (i.e., geodetic) satellites, rocket bodies of the ATLAS family and Jason-1 and Jason-2. TOPEX/Poseidon is the only individual object in this task. This compound solution explores the classifier's ability to deal with higher variability within a class. Further, it allowed to include objects into the pool that formerly had been excluded due to their low number of contributing LCs while still preserving the information about \textit{what} makes objects in a class similar to each other (in this case, first and foremost, geometrical shape). This subset is summarized in Table \ref{tab:onthology_families}.
\begin{table}[ht!]
\caption{\textit{Families} subset (5 classes) with 752 LCs total.}
\centering
\begin{tabular}{|c|c|c|c|c|}
\hline
\textbf{Description}&\textbf{Label}  & \textbf{Unique Objects} & \textbf{LCs} & \textbf{\% of subset}\\
\hline
Jason-1/-2 & JASON & 2 & 240 & 31.9\\
GLONASS & GLONASS & 9 & 177 & 23.5\\
Geodetic Satellites & SPHERE & 4 & 146 & 19.4\\
TOPEX/Poseidon & TOPEX & 1 & 132 & 17.6\\
ATLAS 5 Rocket Bodies & ATLAS & 6 & 57 & 7.6\\
\hline
\end{tabular}
\label{tab:onthology_families}
\end{table}

\textbf{Types:} We recognize that both of the aforementioned labelling schemes could lead to overinterpretation of classifier performance due to the homogeneity within classes. To address this issue, we introduce the third scheme employed in this study, which uses three classes and poses the greatest level of abstraction – active satellites (SAT), rocket bodies (R/B) and spherical satellites (SPH). Inactive satellites as well as debris (both classes that could fit within the logic of this scheme) were omitted because the ensemble sizes introduced severe class imbalances to the task. Hence, this scheme allows to investigate classifier performance without the presence of the most prominent space debris objects (i.e. Jason-2 and TOPEX/Poseidon, both inactive satellites). The labelling has been conducted manually by consulting publicly available information about the objects through their NORAD-ID. Information on this subset can be taken from Table \ref{tab:onthology_types}.
\begin{table}[ht!]
\caption{\textit{Types} subset (3 classes) with 510 LCs total.}
\centering
\begin{tabular}{|c|c|c|c|c|}
\hline
\textbf{Description } &\textbf{Label}  & \textbf{Unique Objects} & \textbf{LCs} &\textbf{\% of subset}\\
\hline
Rocket Bodies    &R/B & 73 & 222 & 43.5\\
Geodetic Satellites & SPH & 4 & 146 & 28.6\\
Active Satellites & SAT &  20 & 142 & 27.8\\
\hline
\end{tabular}
\label{tab:onthology_types}
\end{table}
\section{Methods}
\label{sec:TSC}
\label{sec: methods}
The section introduces the necessary background for feature-based and distance-based classification schemes (see Figure \ref{fig:Methods_workflow}).

\textbf{Mathematical construction of the time series classification problem:} A time series (TS) is a sequence of $n$ ordered real-valued data points $T_{i} = \{t_{1i}, t_{2i}, \dots, t_{ni} \}$. For a given set $T = \{T_{1}, T_{2}, \dots, T_{M} \}$ and a set of associated class labels $ c =\{c_{1}, c_{2}, \dots, c_{M}\}$, Time Series Classification (TSC) aims to find a function that relates the set of time series to the classes \citep{bostrom2015}. 
A particularity of TS is their temporal nature, meaning the order in which attributes are passed to a classifier is relevant. Not all ML models can take this fact into account by default. This has motivated researchers to design algorithms that are specifically tailored towards TSC.   

The main reason for our inclusion of traditional TSC methods in this study is the small quantity of data at our disposal, which limits the meaningfulness of any results achieved with ML models. TSC algorithms offer additional advantages: they are easier to train since they rarely require hyper-parameters and allow more interpretability than NN models, which usually operate as a black box.
TSC approaches can be categorized into distance- and feature-based methods \citep{LeGuennec2016} (applied in Case III and IV from Figure \ref{fig:Methods_workflow}). In the following, we introduce each approach in more detail.

\subsection{Feature Based Classification}

The feature-based approach aims to obtain a new representation of the TS data by selecting its most representative features, reducing the problem's dimensionality in the process. In this context, a feature mapping $\theta_{\ell}: \mathbb{R}^{N} \rightarrow \mathbb{R}^{\ell}$ captures a relevant characteristic. Simple examples of features are the maximum of a time series or its mean \citep{Christ2016}. From a raw LC $T_{i} = \{t_{1i}, t_{2i}, \dots, t_{ni} \} \in \mathbb{R}^{N}$, a feature vector  $F_{i} = \{f_{1i}, f_{2i}, \dots, f_{\ell i}\} \in \mathbb{R}^{\ell}$ with $\ell \ll N$ is obtained. The newly obtained features are then used as the inputs of a classifier (e.g., RDF, XGB, CNN), thereby allowing to bridge the gap between TSC and traditional classification \citep{Abanda2019}. Naturally, the question arises of what kind of features should be computed and which criteria should be applied to assess their relevance. The approach described above is general; it becomes problem-specific when a feature-extraction algorithm is chosen, which gives specific answers to these questions.

\textbf{Feature extraction pipeline:} This work utilizes the $\texttt{TSFresh}$ \citep{Christ2016} feature extraction pipeline to extract statistical features automatically. The feature mappings applied by $\texttt{TSFresh}$ are listed in the appendix of \citep{Christ2016} and are presented in three groups: features from summary statistics (e.g., mean or variance), characteristics of sample distribution (e.g., absolute energy, number of data points above mean) or features derived from observed dynamics (e.g., fast Fourier transformation coefficient or continuous wavelet transformation coefficient). Some of the presented feature mappings require additional parameters. In these cases, $\texttt{TSFresh}$ \footnote{https://tsfresh.readthedocs.io/en/latest/index.html} pipeline computes multiple outputs for the same feature mapping using different parameter options from a $\texttt{feature\_extraction}$ module. 
The \texttt{TSFresh} pipeline utilizes hypothesis tests to evaluate each obtained feature vector, and a vector of p-values is assigned to quantify the significance of each feature. Based on the p-values, \texttt{TSFresh} does not only calculate but also select features for the user. 

\subsection{Distance-Based Classification}
\label{sec:distance_based}
\begin{figure}[h!]
\centering
\includegraphics[width=0.6\textwidth]{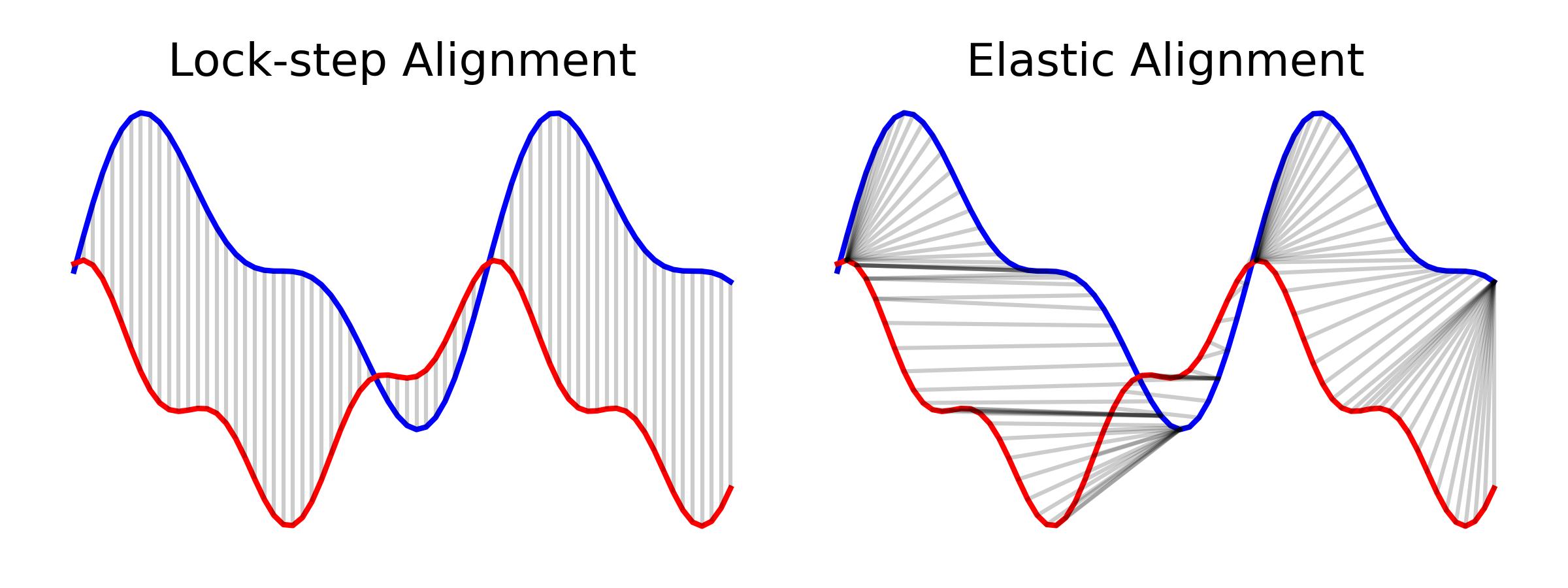}
\caption{Mapping examples of a lock-step measure (e.g., Euclidean Distance) versus an elastic measure (e.g., Dynamic Time Warping) \citep{Wang2013}, graphic inspired by \cite{Abanda2019}.}
\label{fig:alignment}
\end{figure}

Distance based methods obtain a new representation of the TS by applying a distance measure. A distance measure seeks to quantify the similarity between a pair of TS. The choice of the distance measure has a crucial impact on classifier performance. Therefore, various approaches exist that can capture different aspects of dissimilarity \cite{Abanda2019}. Euclidean Distance (ED) and Dynamic Time Warping (DTW) are most commonly applied in TSC \citep{LeGuennec2016}.
ED is a so-called lock-step measure that conducts one-to-one mapping of the data. This means the $i$-th datapoint of one series is mapped to the $i$-th of the other by calculating the distance in euclidean space between samples. 
DTW on the other hand is a more sophisticated algorithm that performs one-to-many mapping (elastic measure) \citep{Wang2013}. In contrary to ED, DTW is able to account for possible misalignment of curves in the time axes which may cause a slight phase shift \citep{Lines2015}. 
Figure \ref{fig:alignment} illustrates the difference between ED and DTW. While ED on the left hand side maps precisely to the corresponding $i$-th index, DTW performs a one-to-many mapping, assigning the $i$-th point of one TS to a fixed number of points in the second. 

The newly obtained representation can be exploited by a distance based classifier e.g., $\texttt{k-NN}$. Albeit its simplicity, the $\texttt{1-NN}$ classifier ($ \texttt{k} = 1$) has been shown to yield competitive performances in the domain of TSC when combined with DTW \citep{Wang2013}. It should be noted, however, that DTW has several shortcomings, namely its quadratic time complexity in TS length as well as its proneness to noise \citep{Schafer2016}.  

\label{sec:methods}
\section{Results}
\label{sec:results}
\begin{figure}[ht!]
\centering
\includegraphics[width=0.5\textwidth]{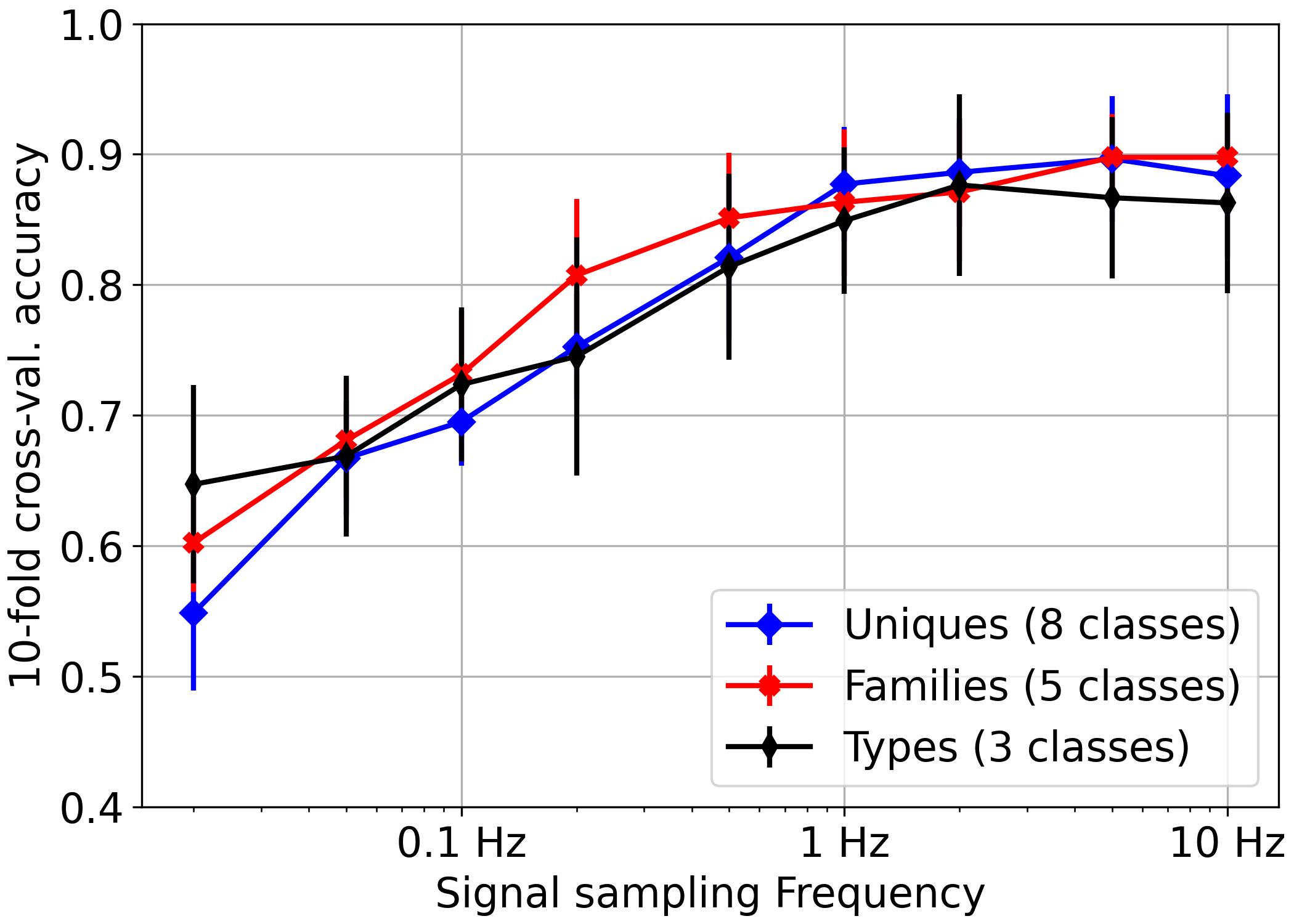}
\caption{Average cross validation performance as a function of sampling frequency for all tasks achieved by the Feautures+XGB classifier.}
\label{fig:downsample}
\end{figure}

\begin{figure}[ht!]
\centering
\begin{subfigure}{0.45\linewidth}
\includegraphics[width=\linewidth]{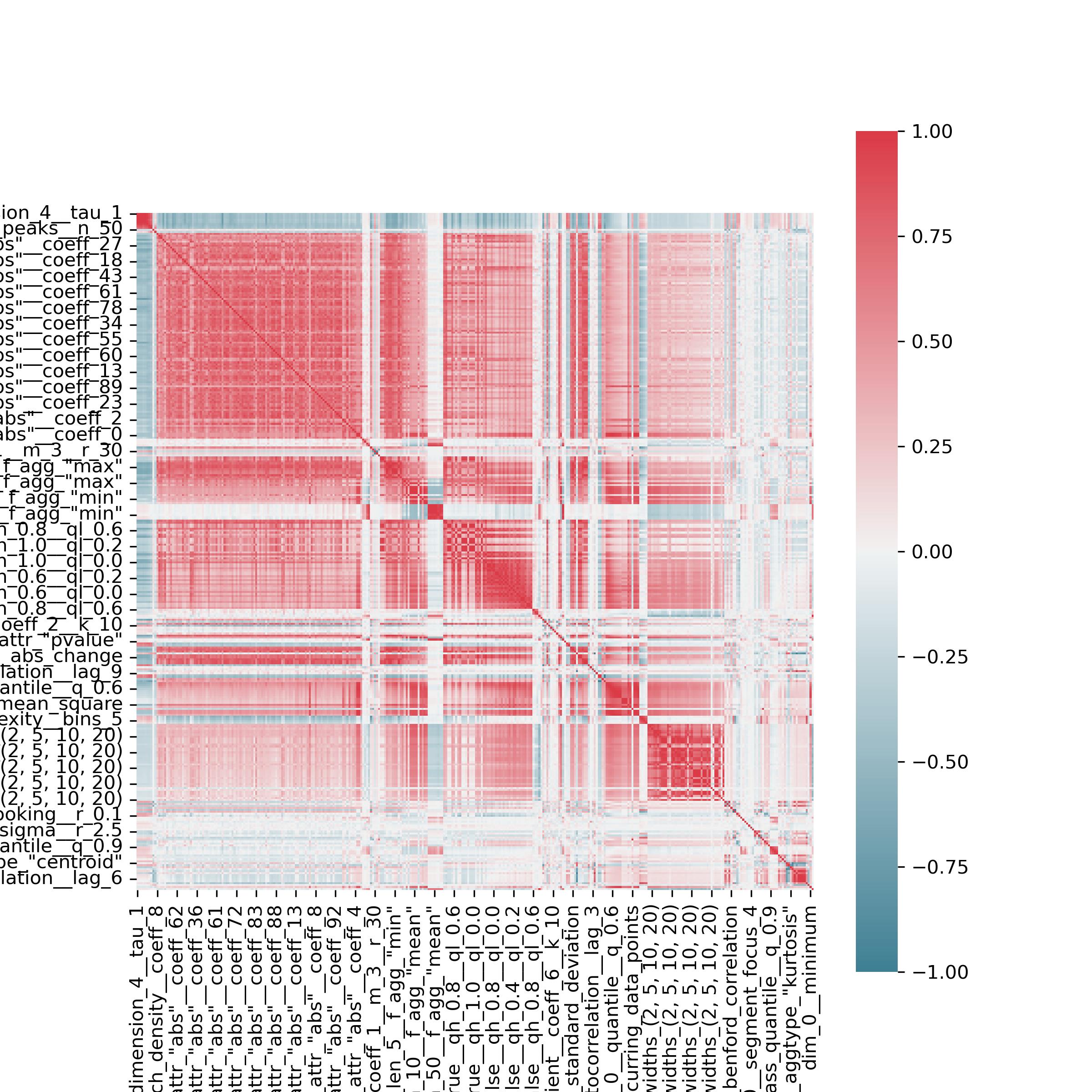}
\caption{}
\label{fig:corr_mat_full}
\end{subfigure}
\begin{subfigure}{0.45\linewidth}
\includegraphics[width=\linewidth]{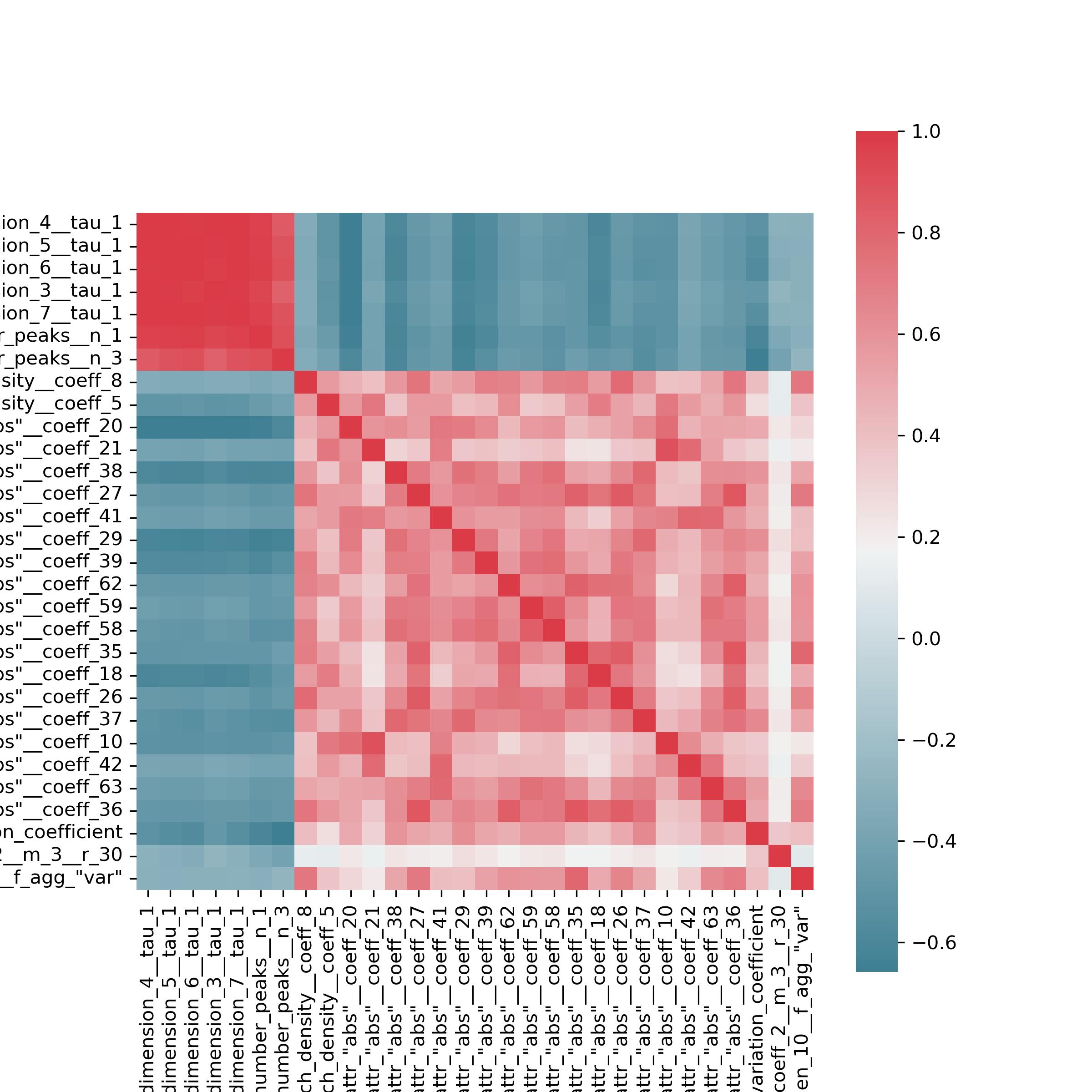}
\caption{}
\label{fig:corr_mat_part}
\end{subfigure}
\caption{(a) The correlation between all the important features (i.e. 342) obtained from $\texttt{TSFresh}$ feature extraction pipeline. (b) The correlation between the first $30$ features has been shown. It is evident that the first seven features are very similar, hence resulting in a high correlation among them.  }
\label{fig:ensemble_hard}
\end{figure}

This section presents our findings regarding the accuracy of all four ML models with LC data and several case studies mentioned in Figure \ref{fig:Methods_workflow}. 
We use decimated LC data to train all ML models. 
We investigated the influence of sampling frequency on model performance by decimating LC data to various sampling frequencies (e.g., $0.1$ Hz, $1$ Hz or $10$ Hz) and training a $\texttt{XGB}$ classifier with extracted features. The accuracy (after 10-fold cross-validation) of the $\texttt{XGB}$ classifier with extracted features from the downsampled versions is depicted in Figure \ref{fig:downsample}. It shows the accuracy of the classifier has been nearly constant from 1 Hz onwards. We have chosen 10 Hz to sample the LC data. 
The choice of length (i.e., 1000) of the decimated LCs was made after careful considerations to minimize information loss, as explained in Section \ref{sec:preprocess}.
Applying the $\texttt{TSFresh}$ feature extraction pipeline to our decimated LC data results in $342$ important features. 
Statistical features could also be extracted from the original LC data using the $\texttt{TSFresh}$ feature extraction pipeline. However, that would increase the computational complexity (in terms of space and time) of running the $\texttt{TSFresh}$ pipeline. The computational complexity of the pipeline heavily depends on the length of the time series. Parallel processing and distributed computation can mitigate this issue. Alternatively, a decimated representation of the time series could reduce the computational burden. We opted for the latter option as, with limited data, distributed computation is not required. Hence, we have explored the feature extraction scheme with decimated LCs. 
To understand the correlation between all 342 features, we compute the correlation between features. Figure \ref{fig:corr_mat_full} shows the correlation between all the statistical features, and Figure \ref{fig:corr_mat_part} shows the first seven features are highly correlated. This leads us to investigate further how many features are crucial to the classification task. 

\begin{figure}
\centering
\begin{subfigure}{0.395\linewidth}
\includegraphics[width=\linewidth]{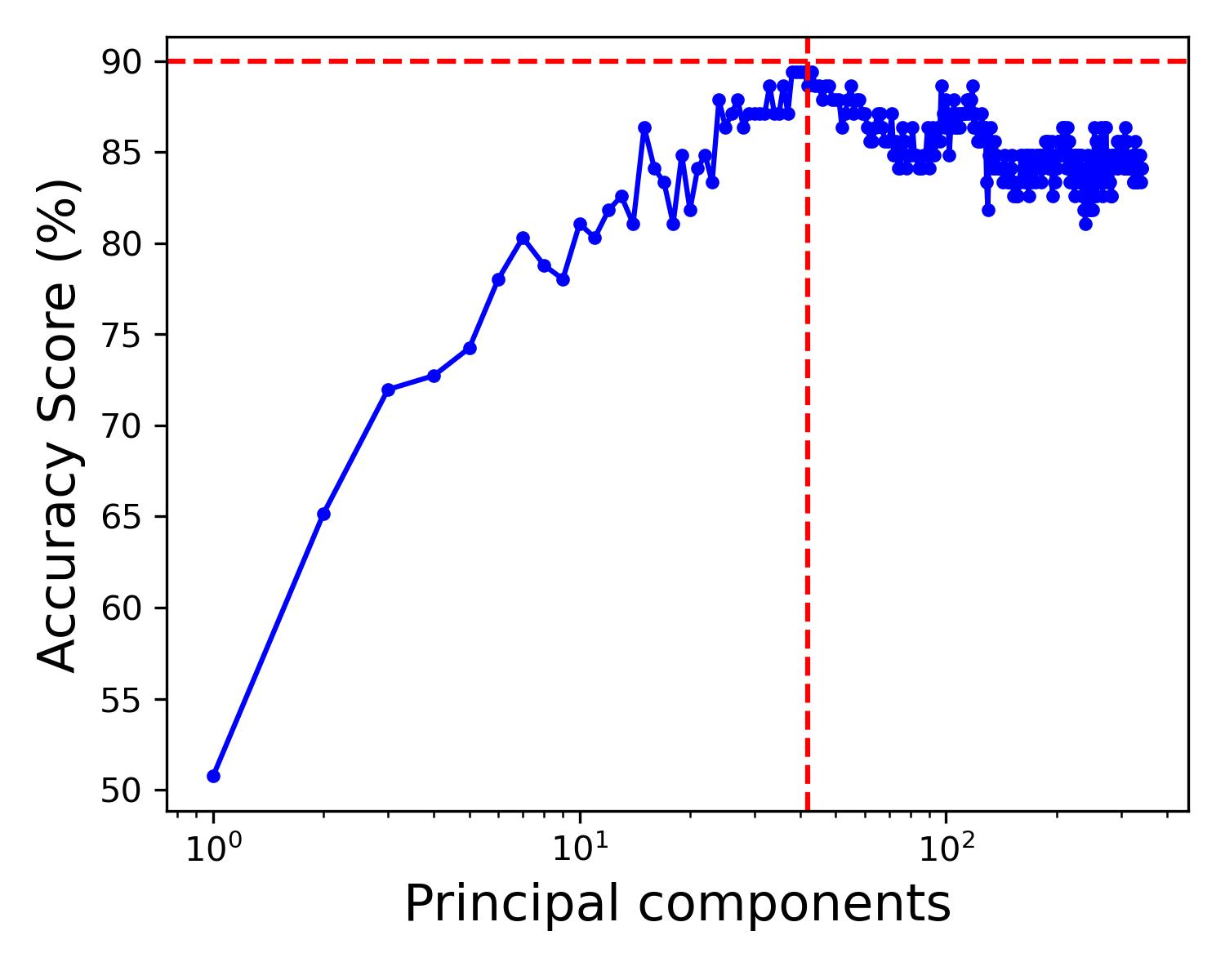}
\caption{}
\label{fig:pca_vs_acc}
\end{subfigure}
\begin{subfigure}{0.45\linewidth}
\includegraphics[width=\linewidth]{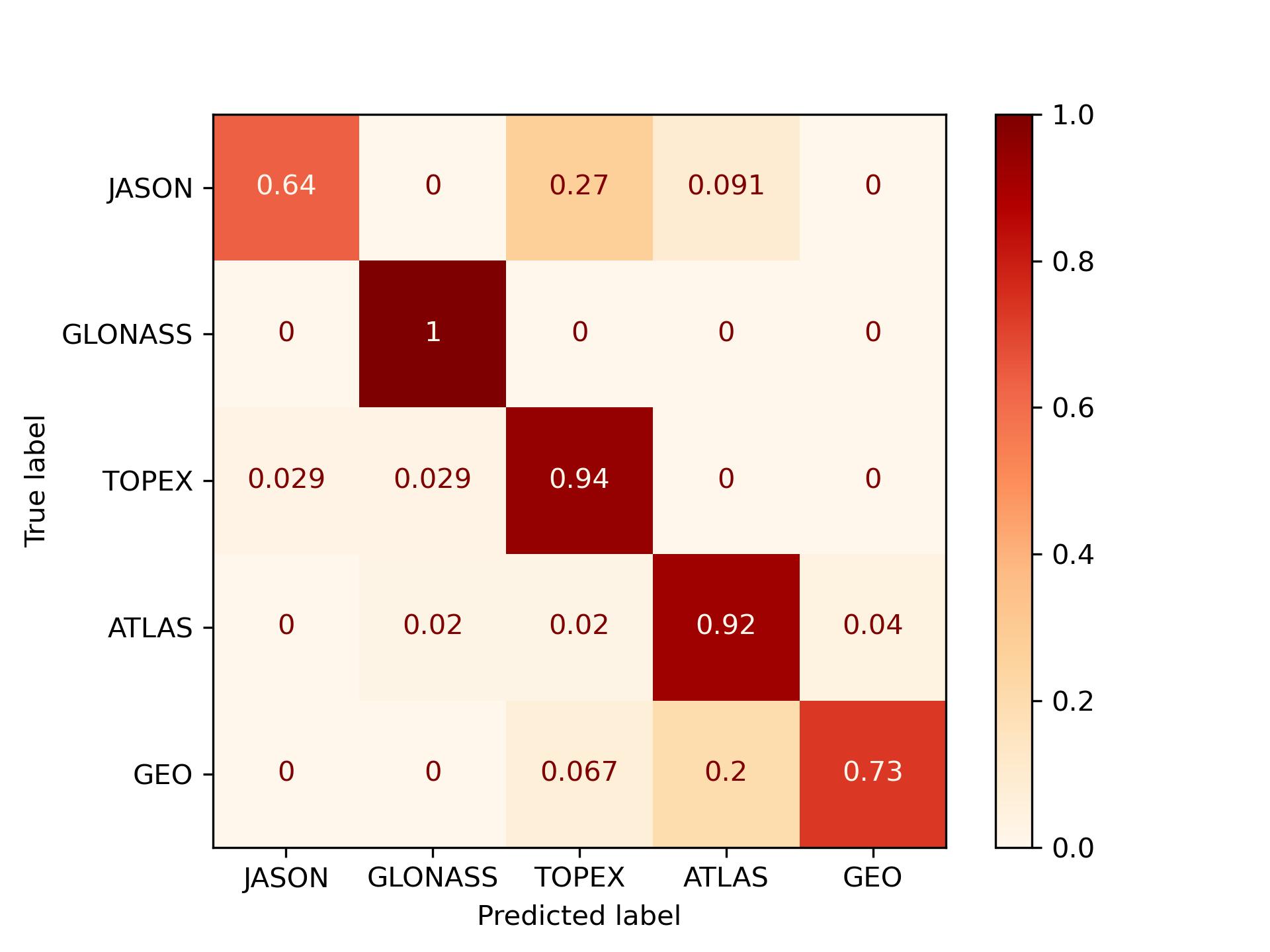}
\caption{}
\label{fig:cm_pca}
\end{subfigure}
\caption{(a) The accuracy score is calculated by varying the principal components of the extracted features ($342$). Top-$42$ principal components are sufficient to represent the maximum variance of the extracted features. (b) The normalized confusion matrix corresponds to the XGB classifier using the top-$42$ principal components of the projected features. The accuracy remains almost identical to that achieved with all 342 features, indicating that some features are redundant and that further feature reduction is possible.}
\label{fig:pca_results}
\end{figure}

\textbf{Feature Selection:} 
To gain insights into the features (obtained from the TSFresh pipeline) used in our study, we investigated them further and found that some features are highly correlated. E.g., the top left corner of the correlation matrix in Figure \ref{fig:ensemble_hard} shows a high correlation between features with a similar name but different numbers at the end. This indicates the feature has been calculated using the same mapping function but with different parameters. Knowing some features are correlated, we applied Principal Components Analysis (PCA) \citep{greenacre2022principal} to further reduce the problem's dimensionality and exclude redundant features. We found the first $42$ components to be relevant for our task. Figure \ref{fig:pca_results} (a) shows that with top-$42$ principal components, we obtained an accuracy score $\sim 90 \%$. 

\subsection{Results on $\texttt{k-NN}$, RDF, XGB}
This section evaluates the performance of the feature-based (Case-III) and distance-based (Case-IV) classification on each of the three subsets (\textit{Uniques}, \textit{Families}, and \textit{Types}). For this, we introduce a total of six model configurations, resulting in eighteen possible combinations of models with tasks. 

We use specific distance metrics for the distance-based method to measure the similarity between the time series (LCs). We decided on DTW as a distance measure since the combination with a $\texttt{k-NN}$ classifier has been shown to yield competitive results in the domain of TSC. In contrast, ED with $\texttt{k-NN}$ is intended to provide a baseline performance (refer to Section \ref{sec:methods}). As suggested by the literature, we choose $\texttt{k} = 1$. The distance-based approach is hence represented by a $\texttt{1-NN}$ classifier in combination with ED (denoted with \texttt{1-NN+ED}) and DTW \texttt{(1-NN+DTW}).

The feature-based method combines the extracted features from the LCs with decision tree classifiers (e.g. RDF, XGB). In the first step, RDF and XGB are each applied directly to LCs, and the results serve as a baseline performance, thereby allowing for comparison to the additional feature extraction step. The notation $\textbf{RDF}$ and $\textbf{XGB}$ is used to define the classification of LCs without extracting features, whereas $\textbf{Features} + \textbf{RDF}$ and $ \textbf{Features} + \textbf{XGB}$ represents the classification after extracting features from LCs using $\texttt{TSFresh}$. We have used the ``Grid Search" scheme to optimize the hyperparameters for RDF and XGB classifiers. For RDF, we found \texttt{criterion = 'entropy', n\_estimators = 200, bootstrap = False, max\_depth = 10, max\_features = 'log2'} to yield the best results. For XGBoost, we obtained \texttt{n\_estimators = 200, objective = 'multi:softprob', tree\_method = 'hist', eta = 0.25, gamma = 0.3, subsample = 0.8, enable\_categorical = True}. It is important to mention that the optimized parameter could vary based on the prior choice of the boundary of the parameters for Grid search. 
\begin{table}[h]
\caption{Average 10-fold cross validation accuracy scores and standard deviation for each classification task and technique achieved on downsampled data. Adding a feature extraction step increases the performance of the \texttt{RDF/XGB} classifiers and enables them to outperform \texttt{1-NN+DTW}.}
\begin{center}
\begin{tabular}{ |c|c|c|c| } 
\hline
\textbf{Classifier} & \textbf{Uniques [\%]} & \textbf{Families [\%]} & \textbf{Types [\%]}\\
\hline
\hline
\texttt{1-NN+ED}	& 48.14 $\pm$5.85 & 53.75$\pm$7.08 &  57.64$\pm$5.56\\ 
\hline
\texttt{1-NN+DTW}	& 76.28$\pm$ 2.68 & 80.32$\pm$4.27 &  81.56$\pm$9.43\\ 
\hline
\hline
\texttt{RDF}		&65.34$\pm$5.34 &68.62$\pm$5.12&  67.45$\pm$7.03\\ 
\hline
\texttt{XGB}		& 58.23$\pm$5.32&63.97$\pm$4.59&  61.37$\pm$7.80\\ 
\hline
\hline
\texttt{Features+RDF} & 87.55$\pm$4.99& 90.70$\pm$5.42& 86.67$\pm$6.43\\
\hline
\texttt{Features+XGB} & 88.17$\pm$4.82 & 89.89$\pm$5.30 &  86.27$\pm$7.44\\
\hline
\end{tabular}
\end{center}
\label{tab:results}
\end{table}

\begin{figure}
\centering
\begin{subfigure}[b]{0.5\textwidth}
\centering
\includegraphics[width=\textwidth]{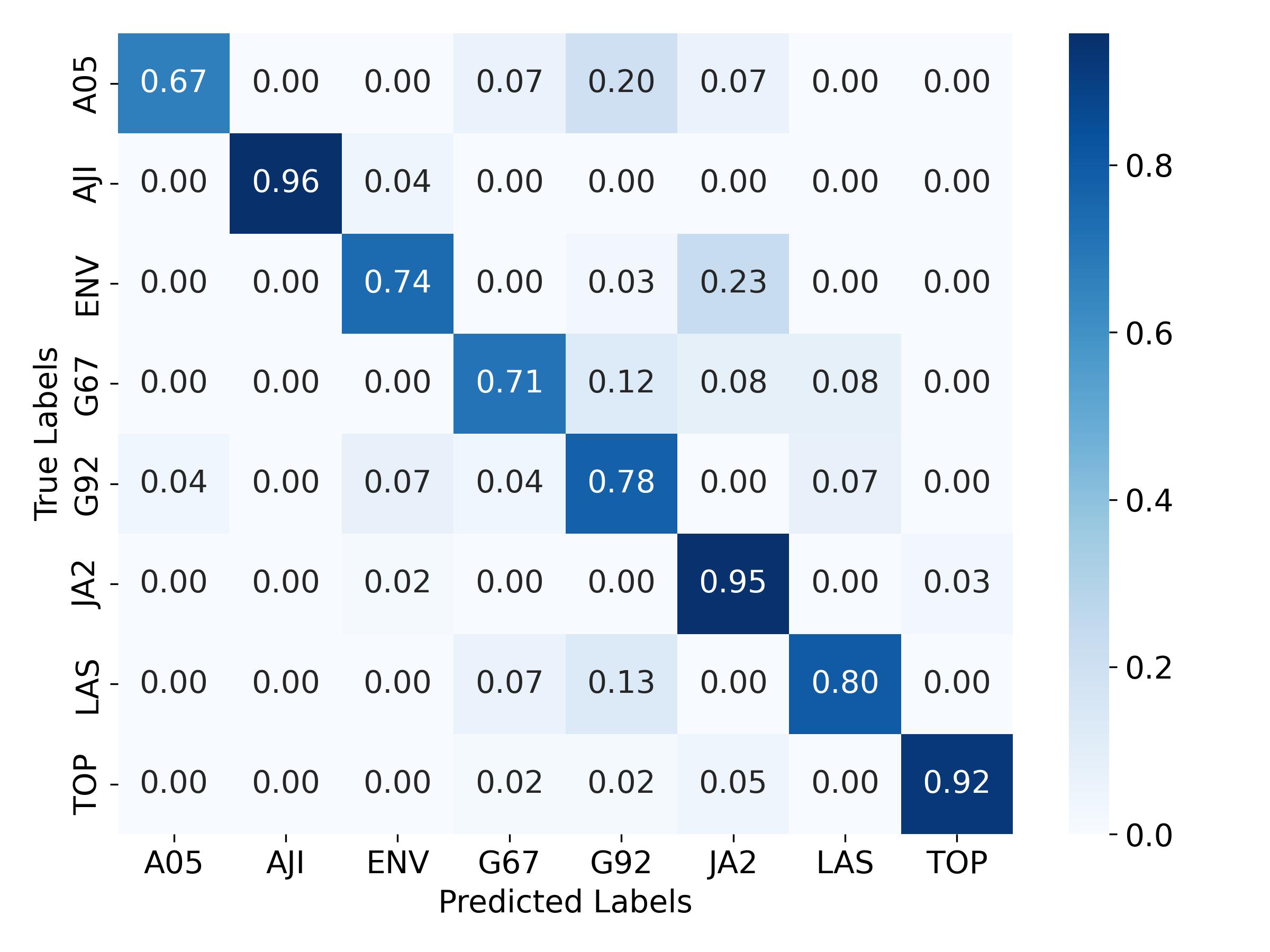}
\caption{\textit{Uniques} Dataset}
\label{fig:cm_a}
\end{subfigure}
\hfill
\begin{subfigure}[b]{0.5\textwidth}
\centering
\includegraphics[width=\textwidth]{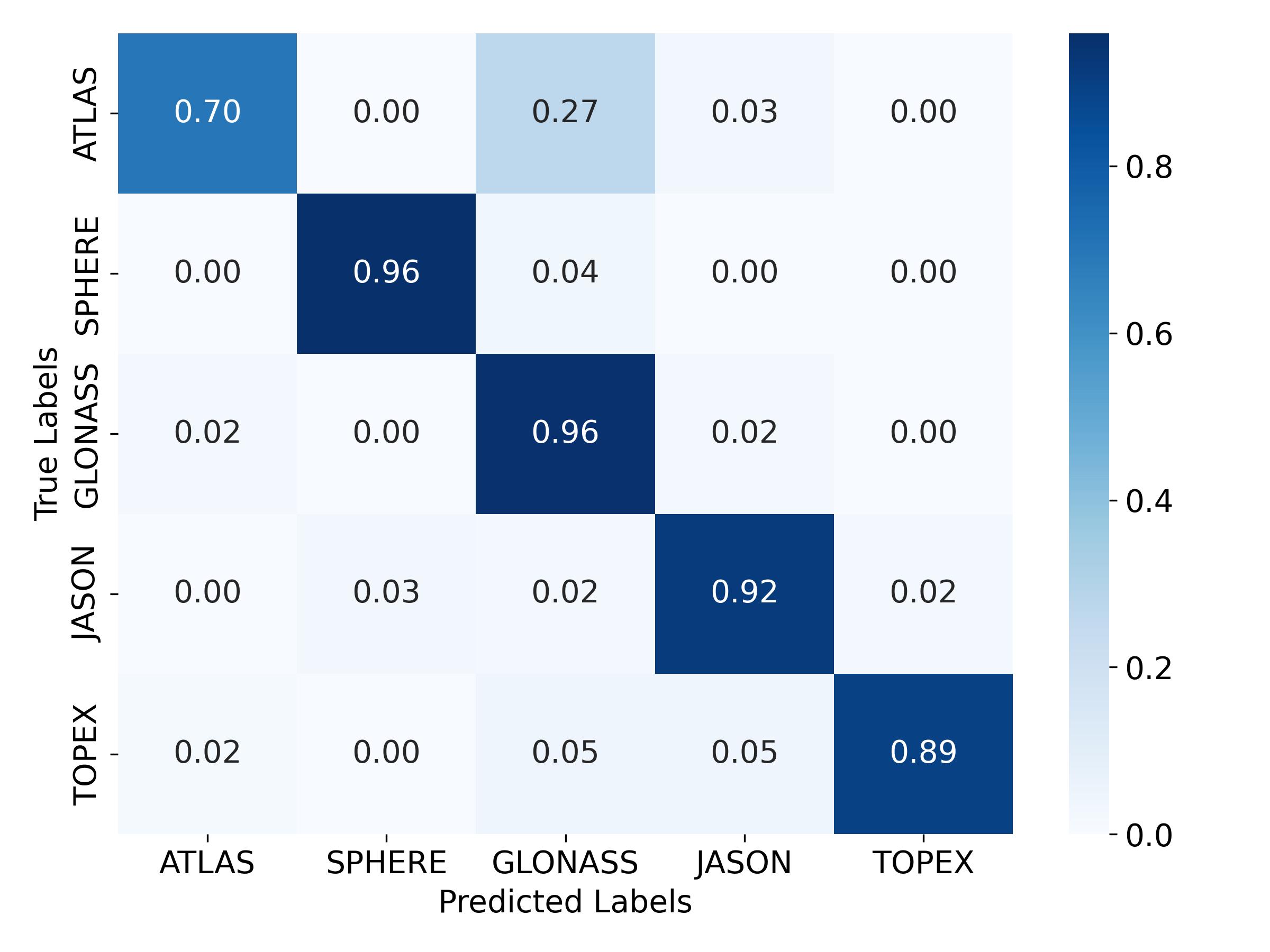}
\caption{\textit{Families} Dataset}
\label{fig:cm_b}
\end{subfigure}
\hfill
\begin{subfigure}[b]{0.5\textwidth}
\centering
\includegraphics[width=\textwidth]{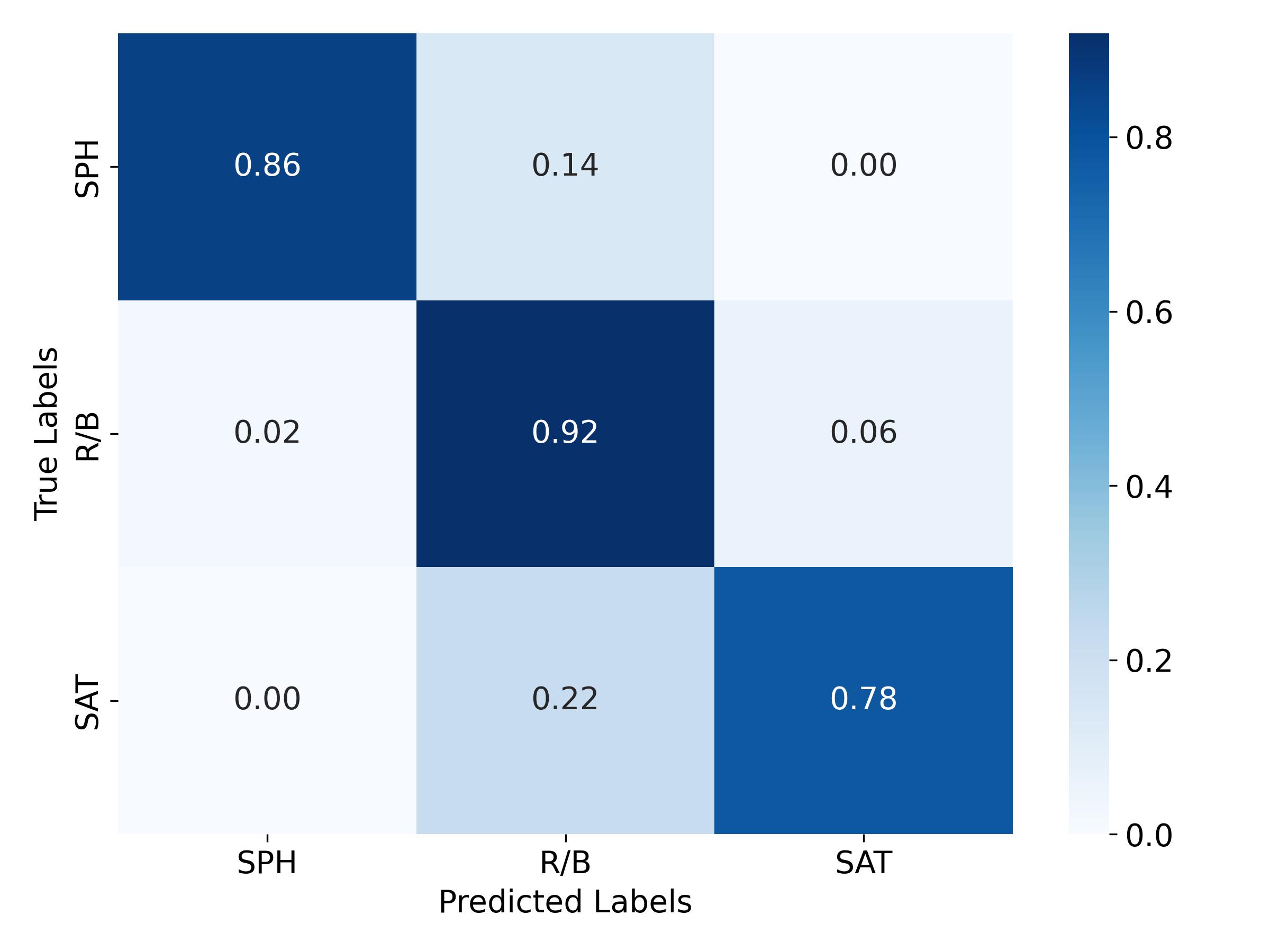}
\caption{\textit{Types} Dataset}
\label{fig:cm_c}
\end{subfigure}
\caption{Normalized confusion matrices for the 10-fold cross validation results achieved by the Features+XGB classifier. Numbers denote average fraction of (mis)classified samples, the diagonals therefore indicate average fraction of correctly classified samples, 1 represents 100\% of LCs in a class.}
\label{fig:cms}
\end{figure}

We assess accuracy using k-fold (here, k = $10$) cross-validation. Table \ref{tab:results} lists the mean accuracies and standard deviations achieved on each model- and task combination. 
The confusion matrices in Fig \ref{fig:cms} show aggregated results on the three tasks (\textit{Uniques, Families, Types}) using the extracted features with $\texttt{XGB}$. 
A number of overall observations can be made from the results listed in Table \ref{tab:results}. As was to be expected, $\texttt{1-NN+DTW}$ outperforms its simpler variant, $\texttt{1-NN+ED}$, $\texttt{RDF}$, and $\texttt{XGB}$. However, a significant increase in performance can be observed for both $\texttt{Features+RDF}$ and $\texttt{Features+XGB}$, further highlighting the importance of the feature extraction step. $\texttt{Features+RDF}$ and $\texttt{Features+XGB}$ yield identical performances, an observation further supported by the similar scores in Table \ref{tab:metrics}. However, when confronted with the vanilla data, $\texttt{RDF}$ outperforms $\texttt{XGB}$. 

\begin{table}[h]
\caption{Average metrics across 10-fold cross validation achieved by the classifiers across three tasks.}
\begin{tabular}{|c|c|c|c|c|c|c|}
\hline
\textbf{Task }& \textbf{Classifier} &\textbf{ Precision} & \textbf{Recall} & \textbf{F1 Score} & \textbf{Specificity} & \textbf{False Positive Rate }\\
\hline
Uniques & \texttt{1-NN+ED} & 0.5609 & 0.4813 & 0.4814 & 0.9258 & 0.0742 \\
\hline 
Uniques & \texttt{1-NN+DTW} & 0.7789 & 0.7628 & 0.7560 & 0.9638 & 0.0362 \\
\hline 
Uniques & \texttt{RDF} & 0.6802 & 0.6534 & 0.6346 & 0.9456 & 0.0544 \\
\hline 
Uniques & \texttt{XGB} & 0.6182 & 0.5823 & 0.5735 & 0.9362 & 0.0638 \\
\hline 
Uniques & \texttt{Features+RDF} & 0.8826 & 0.8755 & 0.8679 & 0.9805 & 0.0195 \\
\hline 
\textbf{Uniques} & \textbf{\texttt{Features+XGB}} & \textbf{0.8890} & \textbf{0.8817} & \textbf{0.8789} & \textbf{0.9819} & \textbf{0.0181} \\
\hline 
\hline
Families & \texttt{1-NN+ED} & 0.5886 & 0.5375 & 0.5349 & 0.8837 & 0.1163 \\
\hline 
Families & \texttt{1-NN+DTW} & 0.8283 & 0.8033 & 0.8031 & 0.9504 & 0.0496 \\
\hline 
Families & \texttt{RDF} & 0.7253 & 0.6862 & 0.6715 & 0.9160 & 0.0840 \\
\hline 
Families & \texttt{XGB} & 0.6471 & 0.6397 & 0.6225 & 0.9048 & 0.0952 \\
\hline 
\textbf{Families }& \textbf{\texttt{Features+RDF}} & \textbf{0.9238} & \textbf{0.9070 }& \textbf{0.9071} & \textbf{0.9756} & \textbf{0.0244} \\
\hline 
Families & \texttt{Features+XGB} & 0.9073 & 0.8989 & 0.8986 & 0.9738 & 0.0262 \\
\hline 
\hline
Types & \texttt{1-NN+ED} & 0.5806 & 0.5765 & 0.5646 & 0.7789 & 0.2211 \\
\hline 
Types & \texttt{1-NN+DTW} & 0.8362 & 0.8157 & 0.8152 & 0.9056 & 0.0944 \\
\hline 
Types & \texttt{RDF} & 0.6819 & 0.6745 & 0.6692 & 0.8328 & 0.1672 \\
\hline 
Types & \texttt{XGB} & 0.6079 & 0.6137 & 0.5991 & 0.7977 & 0.2023 \\
\hline 
\textbf{Types} & \textbf{\texttt{Features+RDF}} & \textbf{ 0.8822} & \textbf{0.8667} & \textbf{0.8650 }& \textbf{0.9282} & \textbf{0.0718} \\
\hline  
Types & \texttt{Features+XGB} & 0.8764 & 0.8627 & 0.8578 & 0.9266 & 0.0734 \\
\hline 
\end{tabular}

\label{tab:metrics}
\end{table}

\vspace{0.25 cm}

In the following, we will discuss results on each dataset in more detail with the aim to infer information about model behavior on class imbalance and variability. 

\textbf{Uniques Dataset:} The \texttt{Features+XGB} approach achieves an overall accuracy of 88.90\% on the \textit{Uniques} dataset (8 classes). Figure \ref{fig:cm_a} depicts the aggregated confusion matrix of the \texttt{Features+XGB} classifier evaluated on this set. From the experiment, it is evident that there is a proneness for ENV to be misclassified as JA2. Both are large, defunct satellites that display visually similar LCs. Further, a certain level of misclassification between G67 and G92 is observed. As both are satellites of the GLONASS constellation and exhibit the same geometry, misclassification is expected. It is, however, compelling to observe that A05 (cylindrical shape), as well as LAS (spherical shape), tend to be mistaken for GLONASS satellites and vice versa. A05 and LAS are minority classes in this task, though the tendency towards GLONASS satellites rather than the dominating class JA2 is somewhat unexpected. On the other hand, the classifier is well able to distinguish JA2, AJI, and TOP.

\begin{figure}
\centering
\includegraphics[width=0.6\textwidth]{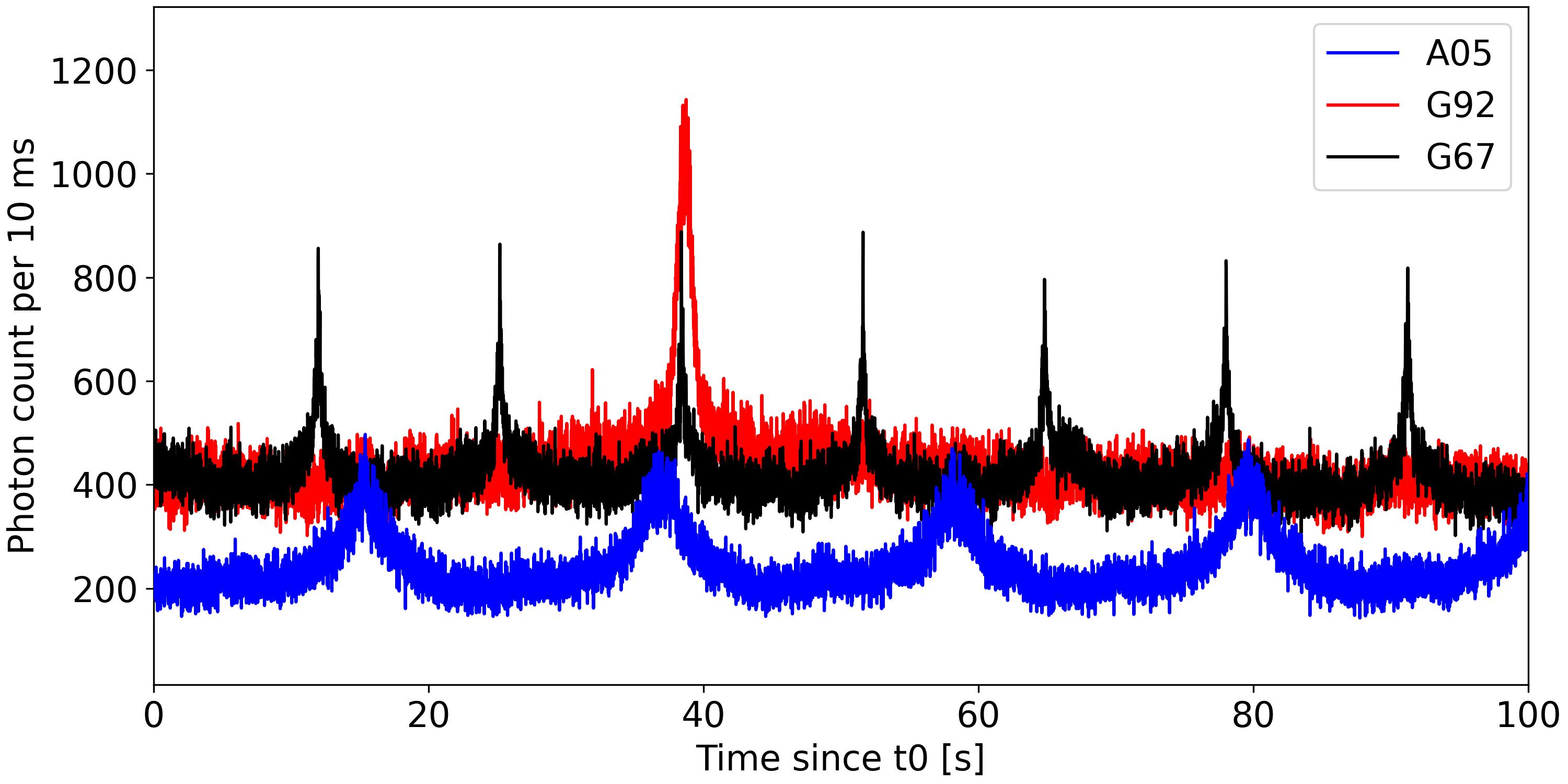}
\caption{Comparison of segments from the objects A05 (class ATLAS) as well as G67 and G92 (class GLONASS). Although of different geometrical type, the lightcurves appear similar.}
\label{fig:comparison}
\end{figure}
\textbf{Families Dataset:} We report an average accuracy of $\sim 90\%$ achieved on the \textit{Families} dataset using \texttt{Features+XGB}. Figure \ref{fig:cm_b} shows the associated confusion matrix. While overall accuracies are high, there is a tendency to misclassify ATLAS objects, a trend already observed in the previous experiment. The class ATLAS comprises six individual rocket bodies of the type ATLAS Centaur. As with the last task, there is a tendency to misclassify these objects as GLONASS satellites, a behavior all high-performing classifiers in this study (\texttt{1-NN+DTW}, \texttt{Features+RDF}, \texttt{Features+XGB}) exhibit. Based on visual inspection of the raw data, as illustrated in Figure \ref{fig:comparison}, we suspect the similarity in signal shape might cause this mistake.

\textbf{Types Dataset:} Extracted \texttt{Features+XGB} achieved an average performance of 86.27\%, Figure \ref{fig:cm_c} depicts the confusion matrix on the \textit{Types} dataset. 
Considering the lower accuracy scores and increased standard deviations, both \texttt{Features+RDF} and \texttt{Features+XGB} models slightly underperformed compared to the other tasks. On average, SAT is wrongly classified as R/B 22\% of the time. Notably, the classes in the task are significantly more heterogeneous than in the other tasks, which could be the reason for obtaining less accuracy.\\

\textbf{Evaluating Train-Test Ratio Effectiveness:} 

\begin{figure}[ht!]
\centering
\includegraphics[width=0.5\textwidth]{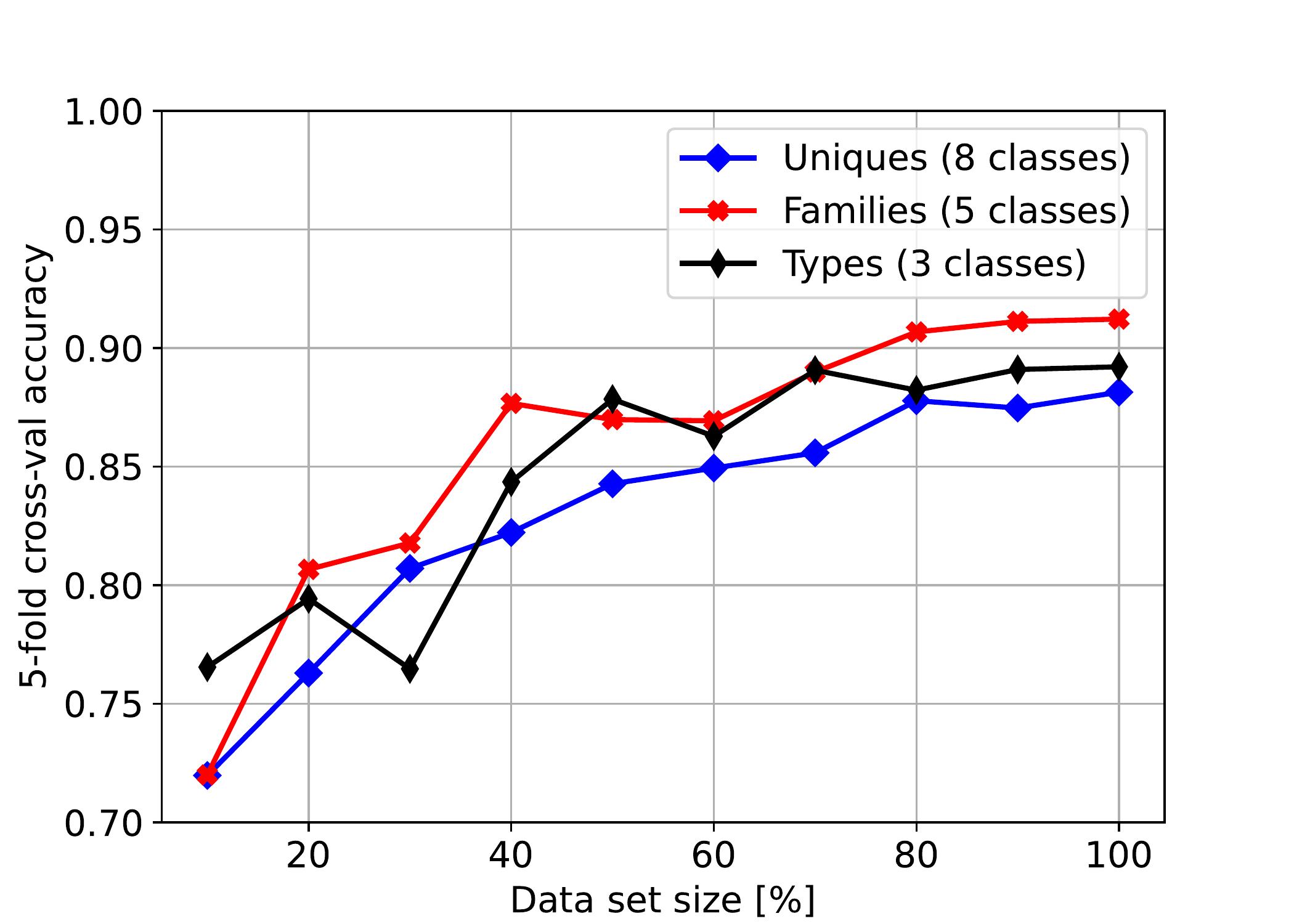}
\caption{Average cross validation performance as a function of data set size for all tasks achieved by the $\texttt{Features+XGB}$ classifier.}
\label{fig:setsize}
\end{figure}

This section evaluates the impact of data quantity on model performance. To do this, we randomly selected various fractions of the total dataset (e.g., 20\%, 30\%) and used them to train and test the model's prediction accuracy, maintaining a fixed 80:20 training-to-testing ratio. Figure \ref{fig:setsize} shows the cross-validation results for the $\texttt{Features+RDF/XGB}$ model as a function of the available data quantity. We used 5-fold cross-validation for this analysis due to the limited dataset sizes. The result indicates that classifier performance across all three tasks begins to plateau around the 80\% data mark, suggesting that the benefit of adding more data starts to diminish beyond this point. This implies that using 64\% of the total data for training, 36\% (with 16\% for validation) for testing is sufficient to assess the performance effectively.
\subsection{Neural Network Classifiers}
\begin{figure}[ht!]
\centering
\includegraphics[width=0.5\textwidth]{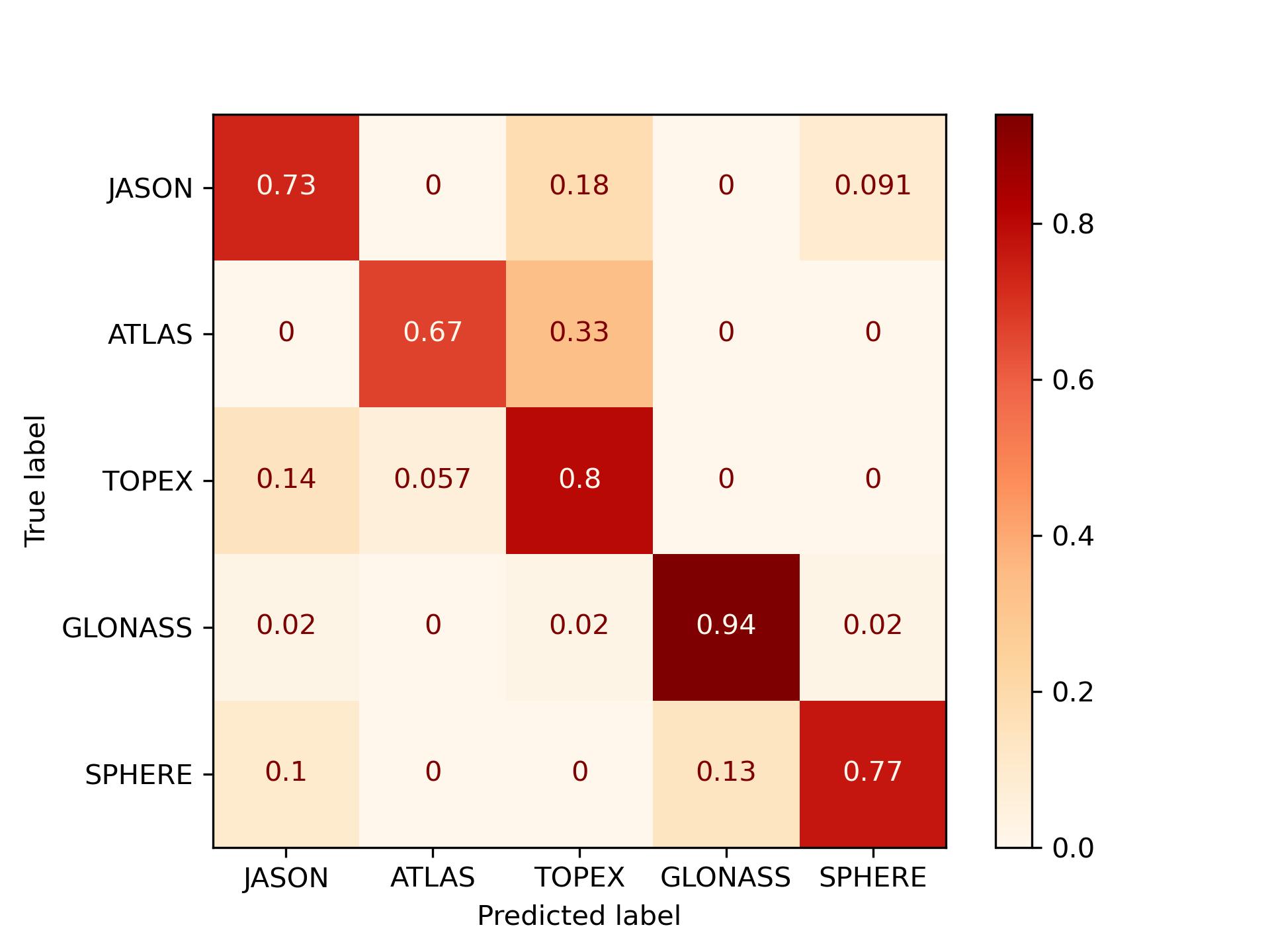}
\caption{The confusion matrix corresponds to the classification of the lightcurves in the original dimensional space. A one-dimensional CNN model has been trained to classify these five families. }
\label{fig:NN_original_dim}
\end{figure}

\begin{figure}[ht!]
\centering
\includegraphics[width=0.5\textwidth]{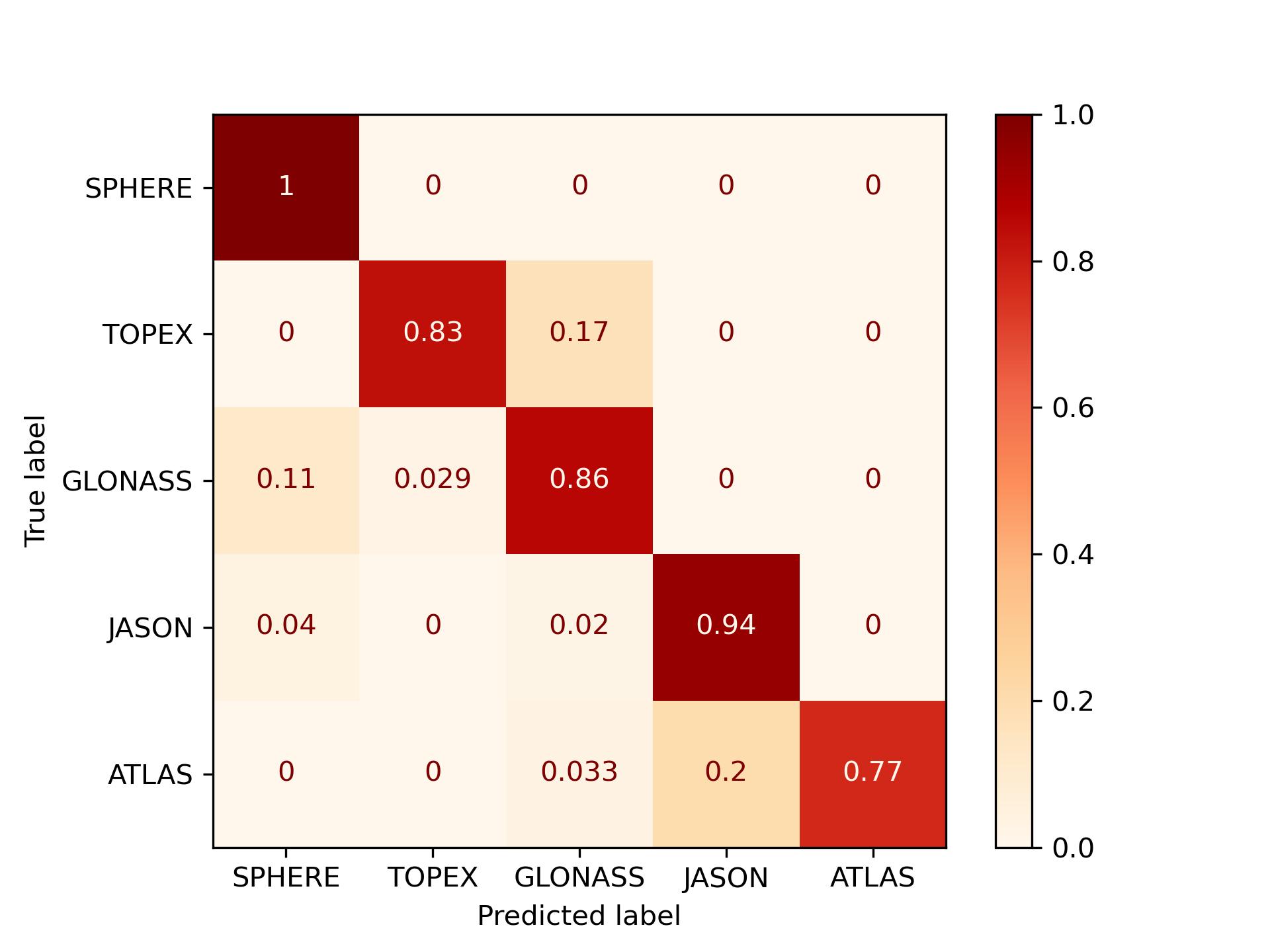}
\caption{The confusion matrix corresponds to the classification of the lightcurves in the down sample space.}
\label{fig:NN_downsample}
\end{figure}

\begin{figure}[ht!]
\centering
\includegraphics[width=0.5\textwidth]{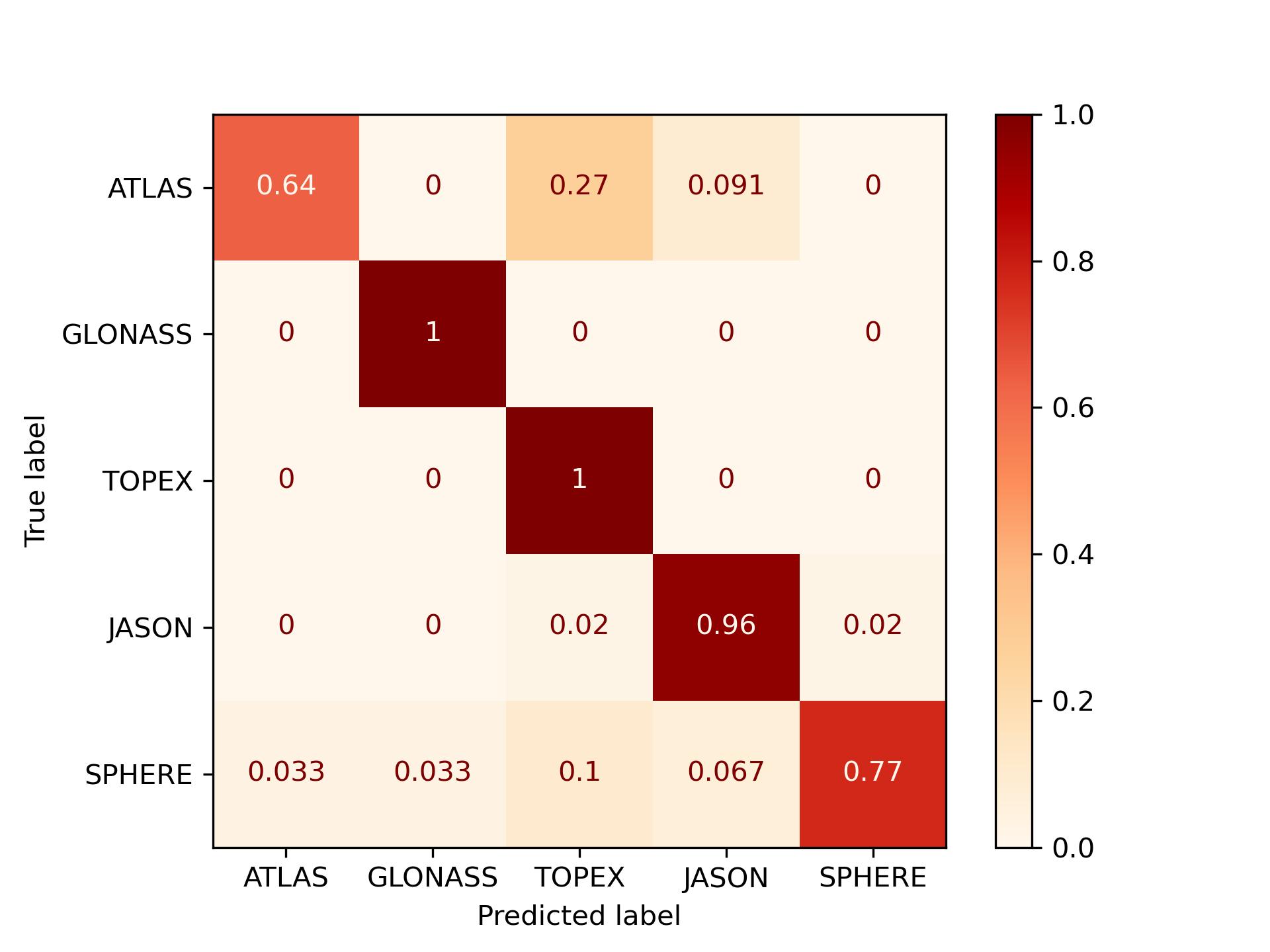}
\caption{The confusion matrix corresponds to the classification of the lightcurves using the extracted features using $\texttt{TSFresh}$. }
\label{fig:NN_feature_ext}
\end{figure}

We explore the Convolutional Neural Network (CNN) to classify light curve data. We perform the training and testing on (i) LCs of original resolution (dimension), (ii) Downsampled LCs, and (iii) Extracted features from LCs. The accuracies for these three cases are $80 \%$, $83 \%$, and $88 \%$, respectively. The confusion matrices for these models are shown in Figures ~\ref{fig:NN_original_dim}, ~\ref{fig:NN_downsample}, ~\ref{fig:NN_feature_ext}. We obtain the highest accuracy with the extracted features that are consistent with other classification models (e.g., Random Forest, Decision Trees). The CNN model could not beat the performance of the xgboost classifier because the model needs more training data. Also, imbalanced classes are another reason for such accuracies. However, for the LCs time series with original resolution (dimension), CNN performs better than other models. This is expected as with the increasing size of the time series; the CNN model could obtain more dynamic features from the data and improve the learning during the training, which is limited to the other classifiers. 

\vspace{0.25cm}
\textbf{Packages used in this study:} The implementation of $\texttt{k-NN}$ and the $\texttt{TSFresh}$ feature extraction step has been done using the AEON-Toolkit\footnote{https://github.com/aeon-toolkit/aeon}. \texttt{TSFresh} and its documentation can be independently used through \footnote{https://tsfresh.readthedocs.io/en/latest/index.html}. For $\texttt{RDF}$, we utilized \texttt{sklearn}-library \cite{pedregosa2011scikit}, $\texttt{XGB}$ is implemented by using the \texttt{xgboost} python library \cite{chen2019package}. For CNN, we used \texttt{Tensorflow} \citep{tensorflow2015-whitepaper}. 
We further utilized \texttt{StratifiedGroupKFold} for stratified split, which prevents $100$-second-segments originating from the same LC, ending up in different sets, and \texttt{GridSearchCV} for hyperparameter tuning. Both are part of the \texttt{sklearn}-library. 

\section{Summary \& Outlook}
\label{sec:conclusion}
This work presents an analysis of the Single Photon LCs collected at Lustbühel Observatory, Graz, Austria, using multiple ML models. 
To the best of our knowledge, this is the first work to demonstrate the analysis of Single Photon LCs with such high accuracy using ML models.
We demonstrate that LCs collected on single photon basis can be utilized to classify space debris using our presented ML formalism with high accuracy. Our results indicate that the automated extracted features with classifiers substantially increase the accuracy of the models, with an average plus in accuracy of $\sim$21\% for RDF and $\sim$27\% for XGB, respectively. Using automated feature extraction in combination with XGB (\texttt{Features+XGB)}, we report an average cross-validated accuracy of $\sim$ 88 \% on \textit{Unqiues}, $\sim$90\% on \textit{Families} and $\sim$86\% on the \textit{Types} dataset. We achieved the highest accuracy by \texttt{Features+RDF} on the \textit{Familes} task with $\sim$ 91 \%. From our results, we conclude that feature extraction is critical to achieving high model performance for space debris classification with ML classifiers, especially for limited data sets with imbalanced classes. Further, methods such as XGB and RDF, in combination with suitable feature extraction, can be a viable alternative to NN, mainly when applied to classification tasks for which well-labeled, high-quality data is difficult to obtain and, therefore, scarce. 

We also performed advanced ML models like CNN for classification and obtained promising results. Due to the limited data size, it could be possible that the CNN model is overfitting the training data. However, the accuracy of the CNN model compares to results presented in the literature \cite{allworth2021, Qashoa2023, Linares2020}. That encourages us to explore advanced NN models in detail for our classification problem with extended data size as a follow-up of this work. 

We recognize that our study has two major limitations. First, we did not use the original resolution of our data due to performance reasons (Case-I, as described in Figure \ref{fig:Methods_workflow}). In the future, we need to explore classification with data of full resolution in order to further profit on potential advantages the high sampling frequency of our setup brings. Second, we ran our experiments on a well-behaved subset of LCs, and while we were able to achieve good results on this subset, being capable of dealing with less than ideal data would allow us to utilize a greater fraction of IWF SPARC. While we are in general able to capture multiple LCs of an object, LC quality will vary with target size and orbit height, and we have to anticipate cases in which we might not be able to acquire a high-quality LC. Therefore, advancing our model's capability to deal with less than ideal data will be of major interest for future work. 

At Lustbühel Observatory, the LC detection package relies on the SLR station's pre-existing infrastructure and tracking capabilities. SPAD detectors for LC measurements can also be incorporated into the routine operation of other SLR stations, further enhancing data quantity. Obtaining more data this way could be cheaper than operating a dedicated telescope or performing a measurement campaign. This opens the option of potential data fusion with the simultaneously conducted SLR measurements. 
It should be mentioned, however, that tying LC measurements to the SLR routine could lead to the overrepresentation of known/cooperative targets in data sets since SLR to space debris is significantly more challenging than to cooperative targets and, therefore, performed with less frequency (cooperative meaning the target carries a retro-reflector). We have the option to avoid this bias by outsourcing the LCs detection package to a second telescope available in our vicinity, thereby achieving independence from the SLR routine. We point out that alternatively, smaller telescopes (e.g., 20 cm) could be sufficient to gather LCs if the tracking capability is given. When discussing the possible compilation of data gathered by multiple devices and/or stations into a single catalogue, the question arises how a ML model's ability to classify data is depended on the used sensor, telescope and site of measurement, for this could introduce unwanted bias in the training process, which is an issue that needs addressing.
Our future research efforts will include investigating the applicability of advanced NN models as more data becomes available over time and exploring data representation. While we strive to enhance the data volume, our efforts might still not be enough to examine the capability of NN models to their full potential, considering out of our raw data (6500 LCs), we chose 852 to be suitable for our experiment, which is $\sim$13\% of the whole catalogue. As demonstrated by, e.g.,\cite{allworth2021}, simulated data could be used to train ML classifiers. We will, therefore, explore options for simulating LCs data while also considering the Single Photon basis of our measurements. 
Also, the follow-up work would explore more complex NN architecture (e.g., Convolutional LSTM, Residual Network) with a large LC dataset. We have already performed Convolutional LSTM with the current dataset and obtained accuracy similar to CNN, but there is a scope to improve the accuracy with the extended dataset. The follow-up paper will report the detailed study of such a model and compare it against the simple three-layer CNN baseline model.

\newpage{}
\section*{References}
\bibliography{references}

\begin{thebibliography}{48}%
\makeatletter
\providecommand \@ifxundefined [1]{%
 \@ifx{#1\undefined}
}%
\providecommand \@ifnum [1]{%
 \ifnum #1\expandafter \@firstoftwo
 \else \expandafter \@secondoftwo
 \fi
}%
\providecommand \@ifx [1]{%
 \ifx #1\expandafter \@firstoftwo
 \else \expandafter \@secondoftwo
 \fi
}%
\providecommand \natexlab [1]{#1}%
\providecommand \enquote  [1]{``#1''}%
\providecommand \bibnamefont  [1]{#1}%
\providecommand \bibfnamefont [1]{#1}%
\providecommand \citenamefont [1]{#1}%
\providecommand \href@noop [0]{\@secondoftwo}%
\providecommand \href [0]{\begingroup \@sanitize@url \@href}%
\providecommand \@href[1]{\@@startlink{#1}\@@href}%
\providecommand \@@href[1]{\endgroup#1\@@endlink}%
\providecommand \@sanitize@url [0]{\catcode `\\12\catcode `\$12\catcode
  `\&12\catcode `\#12\catcode `\^12\catcode `\_12\catcode `\%12\relax}%
\providecommand \@@startlink[1]{}%
\providecommand \@@endlink[0]{}%
\providecommand \url  [0]{\begingroup\@sanitize@url \@url }%
\providecommand \@url [1]{\endgroup\@href {#1}{\urlprefix }}%
\providecommand \urlprefix  [0]{URL }%
\providecommand \Eprint [0]{\href }%
\providecommand \doibase [0]{https://doi.org/}%
\providecommand \selectlanguage [0]{\@gobble}%
\providecommand \bibinfo  [0]{\@secondoftwo}%
\providecommand \bibfield  [0]{\@secondoftwo}%
\providecommand \translation [1]{[#1]}%
\providecommand \BibitemOpen [0]{}%
\providecommand \bibitemStop [0]{}%
\providecommand \bibitemNoStop [0]{.\EOS\space}%
\providecommand \EOS [0]{\spacefactor3000\relax}%
\providecommand \BibitemShut  [1]{\csname bibitem#1\endcsname}%
\let\auto@bib@innerbib\@empty
\bibitem [{ESA({\natexlab{a}})}]{ESAReport2023}%
  \BibitemOpen
  \href
  {https://www.esa.int/Space\_Safety/ESA\_s\_Space\_Environment\_Report\_2023}
  {\bibinfo {title} {The european space agency’s space environment report
  2023}} ({\natexlab{a}}),\ \bibinfo {note} {accessed: 29 August
  2024}\BibitemShut {NoStop}%
\bibitem [{ESA({\natexlab{b}})}]{ESASpaceDebris}%
  \BibitemOpen
  \href
  {https://www.esa.int/Space\_Safety/Space\_Debris/Space\_debris\_by\_the\_numbers}
  {\bibinfo {title} {Space debris by the numbers}} ({\natexlab{b}}),\ \bibinfo
  {note} {accessed: 29 August 2024}\BibitemShut {NoStop}%
\bibitem [{\citenamefont {Schildknecht}(2007)}]{Schildknecht2007}%
  \BibitemOpen
  \bibfield  {author} {\bibinfo {author} {\bibfnamefont {T.}~\bibnamefont
  {Schildknecht}},\ }\bibfield  {title} {\bibinfo {title} {Optical surveys for
  space debris},\ }\href {https://doi.org/10.1007/s00159-006-0003-9} {\bibfield
   {journal} {\bibinfo  {journal} {The Astronomy and Astrophysics Review}\
  }\textbf {\bibinfo {volume} {14}},\ \bibinfo {pages} {41} (\bibinfo {year}
  {2007})}\BibitemShut {NoStop}%
\bibitem [{\citenamefont {Kessler}\ and\ \citenamefont
  {Cour-Palais}(1978)}]{kessler1978}%
  \BibitemOpen
  \bibfield  {author} {\bibinfo {author} {\bibfnamefont {D.~J.}\ \bibnamefont
  {Kessler}}\ and\ \bibinfo {author} {\bibfnamefont {B.~G.}\ \bibnamefont
  {Cour-Palais}},\ }\bibfield  {title} {\bibinfo {title} {Collision frequency
  of artificial satellites: The creation of a debris belt},\ }\href
  {https://doi.org/10.1029/JA083iA06p02637} {\bibfield  {journal} {\bibinfo
  {journal} {Journal of Geophysical Research: Space Physics}\ }\textbf
  {\bibinfo {volume} {83}},\ \bibinfo {pages} {2637} (\bibinfo {year}
  {1978})}\BibitemShut {NoStop}%
\bibitem [{\citenamefont {Fan}\ and\ \citenamefont {Frueh}(2020)}]{Fan2020}%
  \BibitemOpen
  \bibfield  {author} {\bibinfo {author} {\bibfnamefont {S.}~\bibnamefont
  {Fan}}\ and\ \bibinfo {author} {\bibfnamefont {C.}~\bibnamefont {Frueh}},\
  }\bibfield  {title} {\bibinfo {title} {A direct light curve inversion scheme
  in the presence of measurement noise},\ }\href
  {https://doi.org/10.1007/s40295-019-00190-3} {\bibfield  {journal} {\bibinfo
  {journal} {The Journal of the Astronautical Sciences}\ }\textbf {\bibinfo
  {volume} {67}},\ \bibinfo {pages} {740} (\bibinfo {year} {2020})}\BibitemShut
  {NoStop}%
\bibitem [{\citenamefont {{\v{S}}ilha}\ \emph {et~al.}(2021)\citenamefont
  {{\v{S}}ilha}, \citenamefont {Zigo}, \citenamefont {Hrob{{\'a}}r},
  \citenamefont {Jev{\v{c}}{{\'a}}k},\ and\ \citenamefont
  {Vere{\v{s}}v{{\'a}}rska}}]{Silha2021}%
  \BibitemOpen
  \bibfield  {author} {\bibinfo {author} {\bibfnamefont {J.}~\bibnamefont
  {{\v{S}}ilha}}, \bibinfo {author} {\bibfnamefont {M.}~\bibnamefont {Zigo}},
  \bibinfo {author} {\bibfnamefont {T.}~\bibnamefont {Hrob{{\'a}}r}}, \bibinfo
  {author} {\bibfnamefont {P.}~\bibnamefont {Jev{\v{c}}{{\'a}}k}},\ and\
  \bibinfo {author} {\bibfnamefont {M.}~\bibnamefont
  {Vere{\v{s}}v{{\'a}}rska}},\ }\bibfield  {title} {\bibinfo {title} {Light
  curves application to space debris characterization and classification},\
  }\href
  {https://doi.org/https://conference.sdo.esoc.esa.int/proceedings/sdc8/paper/168/SDC8-paper168.pdf}
  {\bibfield  {journal} {\bibinfo  {journal} {complexity}\ }\textbf {\bibinfo
  {volume} {10}},\ \bibinfo {pages} {3} (\bibinfo {year} {2021})}\BibitemShut
  {NoStop}%
\bibitem [{\citenamefont {Kucharski}\ \emph {et~al.}(2021)\citenamefont
  {Kucharski}, \citenamefont {Kirchner}, \citenamefont {Jah}, \citenamefont
  {Bennett}, \citenamefont {Koidl}, \citenamefont {Steindorfer},\ and\
  \citenamefont {Wang}}]{Kucharski2021}%
  \BibitemOpen
  \bibfield  {author} {\bibinfo {author} {\bibfnamefont {D.}~\bibnamefont
  {Kucharski}}, \bibinfo {author} {\bibfnamefont {G.}~\bibnamefont {Kirchner}},
  \bibinfo {author} {\bibfnamefont {M.~K.}\ \bibnamefont {Jah}}, \bibinfo
  {author} {\bibfnamefont {J.~C.}\ \bibnamefont {Bennett}}, \bibinfo {author}
  {\bibfnamefont {F.}~\bibnamefont {Koidl}}, \bibinfo {author} {\bibfnamefont
  {M.~A.}\ \bibnamefont {Steindorfer}},\ and\ \bibinfo {author} {\bibfnamefont
  {P.}~\bibnamefont {Wang}},\ }\bibfield  {title} {\bibinfo {title} {Full
  attitude state reconstruction of tumbling space debris topex/poseidon via
  light-curve inversion with quanta photogrammetry},\ }\href
  {https://doi.org/10.1016/j.actaastro.2021.06.032} {\bibfield  {journal}
  {\bibinfo  {journal} {Acta Astronautica}\ }\textbf {\bibinfo {volume}
  {187}},\ \bibinfo {pages} {115} (\bibinfo {year} {2021})}\BibitemShut
  {NoStop}%
\bibitem [{\citenamefont {{\v{S}}ilha}\ \emph {et~al.}(2018)\citenamefont
  {{\v{S}}ilha}, \citenamefont {Pittet}, \citenamefont {Hamara},\ and\
  \citenamefont {Schildknecht}}]{Silha2018}%
  \BibitemOpen
  \bibfield  {author} {\bibinfo {author} {\bibfnamefont {J.}~\bibnamefont
  {{\v{S}}ilha}}, \bibinfo {author} {\bibfnamefont {J.-N.}\ \bibnamefont
  {Pittet}}, \bibinfo {author} {\bibfnamefont {M.}~\bibnamefont {Hamara}},\
  and\ \bibinfo {author} {\bibfnamefont {T.}~\bibnamefont {Schildknecht}},\
  }\bibfield  {title} {\bibinfo {title} {Apparent rotation properties of space
  debris extracted from photometric measurements},\ }\href
  {https://doi.org/10.1016/j.asr.2017.10.048} {\bibfield  {journal} {\bibinfo
  {journal} {Advances in space research}\ }\textbf {\bibinfo {volume} {61}},\
  \bibinfo {pages} {844} (\bibinfo {year} {2018})}\BibitemShut {NoStop}%
\bibitem [{\citenamefont {Allworth}\ \emph {et~al.}(2021)\citenamefont
  {Allworth}, \citenamefont {Windrim}, \citenamefont {Bennett},\ and\
  \citenamefont {Bryson}}]{allworth2021}%
  \BibitemOpen
  \bibfield  {author} {\bibinfo {author} {\bibfnamefont {J.}~\bibnamefont
  {Allworth}}, \bibinfo {author} {\bibfnamefont {L.}~\bibnamefont {Windrim}},
  \bibinfo {author} {\bibfnamefont {J.}~\bibnamefont {Bennett}},\ and\ \bibinfo
  {author} {\bibfnamefont {M.}~\bibnamefont {Bryson}},\ }\bibfield  {title}
  {\bibinfo {title} {A transfer learning approach to space debris
  classification using observational light curve data},\ }\href
  {https://doi.org/10.1016/j.actaastro.2021.01.048} {\bibfield  {journal}
  {\bibinfo  {journal} {Acta Astronautica}\ }\textbf {\bibinfo {volume}
  {181}},\ \bibinfo {pages} {301} (\bibinfo {year} {2021})}\BibitemShut
  {NoStop}%
\bibitem [{\citenamefont {Burton}\ and\ \citenamefont
  {Frueh}(2021)}]{burton2021}%
  \BibitemOpen
  \bibfield  {author} {\bibinfo {author} {\bibfnamefont {A.}~\bibnamefont
  {Burton}}\ and\ \bibinfo {author} {\bibfnamefont {C.}~\bibnamefont {Frueh}},\
  }\bibfield  {title} {\bibinfo {title} {Two methods for light curve inversion
  for space object attitude determination},\ }in\ \href
  {https://doi.org/https://conference.sdo.esoc.esa.int/proceedings/sdc8/paper/18/SDC8-paper18.pdf}
  {\emph {\bibinfo {booktitle} {Proceedings of the 8th European conference on
  space debris}}}\ (\bibinfo {year} {2021})\BibitemShut {NoStop}%
\bibitem [{\citenamefont {Linares}\ \emph {et~al.}(2020)\citenamefont
  {Linares}, \citenamefont {Furfaro},\ and\ \citenamefont
  {Reddy}}]{Linares2020}%
  \BibitemOpen
  \bibfield  {author} {\bibinfo {author} {\bibfnamefont {R.}~\bibnamefont
  {Linares}}, \bibinfo {author} {\bibfnamefont {R.}~\bibnamefont {Furfaro}},\
  and\ \bibinfo {author} {\bibfnamefont {V.}~\bibnamefont {Reddy}},\ }\bibfield
   {title} {\bibinfo {title} {Space objects classification via light-curve
  measurements using deep convolutional neural networks},\ }\href
  {https://doi.org/10.1007/s40295-019-00208-w} {\bibfield  {journal} {\bibinfo
  {journal} {The Journal of the Astronautical Sciences}\ }\textbf {\bibinfo
  {volume} {67}},\ \bibinfo {pages} {1063} (\bibinfo {year}
  {2020})}\BibitemShut {NoStop}%
\bibitem [{\citenamefont {Wu}\ and\ \citenamefont
  {Feng}(2018)}]{wu2018development}%
  \BibitemOpen
  \bibfield  {author} {\bibinfo {author} {\bibfnamefont {Y.-c.}\ \bibnamefont
  {Wu}}\ and\ \bibinfo {author} {\bibfnamefont {J.-w.}\ \bibnamefont {Feng}},\
  }\bibfield  {title} {\bibinfo {title} {Development and application of
  artificial neural network},\ }\href
  {https://doi.org/10.1007/s11277-017-5224-x} {\bibfield  {journal} {\bibinfo
  {journal} {Wireless Personal Communications}\ }\textbf {\bibinfo {volume}
  {102}},\ \bibinfo {pages} {1645} (\bibinfo {year} {2018})}\BibitemShut
  {NoStop}%
\bibitem [{\citenamefont {Albawi}\ \emph {et~al.}(2017)\citenamefont {Albawi},
  \citenamefont {Mohammed},\ and\ \citenamefont
  {Al-Zawi}}]{albawi2017understanding}%
  \BibitemOpen
  \bibfield  {author} {\bibinfo {author} {\bibfnamefont {S.}~\bibnamefont
  {Albawi}}, \bibinfo {author} {\bibfnamefont {T.~A.}\ \bibnamefont
  {Mohammed}},\ and\ \bibinfo {author} {\bibfnamefont {S.}~\bibnamefont
  {Al-Zawi}},\ }\bibfield  {title} {\bibinfo {title} {Understanding of a
  convolutional neural network},\ }in\ \href
  {https://doi.org/10.1109/ICEngTechnol.2017.8308186} {\emph {\bibinfo
  {booktitle} {2017 international conference on engineering and technology
  (ICET)}}}\ (\bibinfo {organization} {Ieee},\ \bibinfo {year} {2017})\ pp.\
  \bibinfo {pages} {1--6}\BibitemShut {NoStop}%
\bibitem [{\citenamefont {Wu}(2017)}]{wu2017introduction}%
  \BibitemOpen
  \bibfield  {author} {\bibinfo {author} {\bibfnamefont {J.}~\bibnamefont
  {Wu}},\ }\bibfield  {title} {\bibinfo {title} {Introduction to convolutional
  neural networks},\ }\href
  {https://doi.org/https://cs.nju.edu.cn/wujx/paper/CNN.pdf} {\bibfield
  {journal} {\bibinfo  {journal} {National Key Lab for Novel Software
  Technology. Nanjing University. China}\ }\textbf {\bibinfo {volume} {5}},\
  \bibinfo {pages} {495} (\bibinfo {year} {2017})}\BibitemShut {NoStop}%
\bibitem [{\citenamefont {Gu}\ \emph {et~al.}(2018)\citenamefont {Gu},
  \citenamefont {Wang}, \citenamefont {Kuen}, \citenamefont {Ma}, \citenamefont
  {Shahroudy}, \citenamefont {Shuai}, \citenamefont {Liu}, \citenamefont
  {Wang}, \citenamefont {Wang}, \citenamefont {Cai} \emph
  {et~al.}}]{gu2018recent}%
  \BibitemOpen
  \bibfield  {author} {\bibinfo {author} {\bibfnamefont {J.}~\bibnamefont
  {Gu}}, \bibinfo {author} {\bibfnamefont {Z.}~\bibnamefont {Wang}}, \bibinfo
  {author} {\bibfnamefont {J.}~\bibnamefont {Kuen}}, \bibinfo {author}
  {\bibfnamefont {L.}~\bibnamefont {Ma}}, \bibinfo {author} {\bibfnamefont
  {A.}~\bibnamefont {Shahroudy}}, \bibinfo {author} {\bibfnamefont
  {B.}~\bibnamefont {Shuai}}, \bibinfo {author} {\bibfnamefont
  {T.}~\bibnamefont {Liu}}, \bibinfo {author} {\bibfnamefont {X.}~\bibnamefont
  {Wang}}, \bibinfo {author} {\bibfnamefont {G.}~\bibnamefont {Wang}}, \bibinfo
  {author} {\bibfnamefont {J.}~\bibnamefont {Cai}}, \emph {et~al.},\ }\bibfield
   {title} {\bibinfo {title} {Recent advances in convolutional neural
  networks},\ }\href {https://doi.org/10.1016/j.patcog.2017.10.013} {\bibfield
  {journal} {\bibinfo  {journal} {Pattern recognition}\ }\textbf {\bibinfo
  {volume} {77}},\ \bibinfo {pages} {354} (\bibinfo {year} {2018})}\BibitemShut
  {NoStop}%
\bibitem [{\citenamefont {Furfaro}\ \emph {et~al.}(2016)\citenamefont
  {Furfaro}, \citenamefont {Linares}, \citenamefont {Gaylor}, \citenamefont
  {Jah},\ and\ \citenamefont {Walls}}]{Furfaro2016}%
  \BibitemOpen
  \bibfield  {author} {\bibinfo {author} {\bibfnamefont {R.}~\bibnamefont
  {Furfaro}}, \bibinfo {author} {\bibfnamefont {R.}~\bibnamefont {Linares}},
  \bibinfo {author} {\bibfnamefont {D.}~\bibnamefont {Gaylor}}, \bibinfo
  {author} {\bibfnamefont {M.}~\bibnamefont {Jah}},\ and\ \bibinfo {author}
  {\bibfnamefont {R.}~\bibnamefont {Walls}},\ }\bibfield  {title} {\bibinfo
  {title} {Resident space object characterization and behavior understanding
  via machine learning and ontology-based bayesian networks},\ }in\ \href
  {https://doi.org/https://amostech.com/TechnicalPapers/2016/SSA-Algorithms/Furfaro.pdf}
  {\emph {\bibinfo {booktitle} {Advanced Maui Optical and Space Surveillance
  Technologies Conference}}}\ (\bibinfo {year} {2016})\ p.~\bibinfo {pages}
  {35}\BibitemShut {NoStop}%
\bibitem [{\citenamefont {Furfaro}\ \emph {et~al.}(2019)\citenamefont
  {Furfaro}, \citenamefont {Linares},\ and\ \citenamefont
  {Reddy}}]{Furfaro2019}%
  \BibitemOpen
  \bibfield  {author} {\bibinfo {author} {\bibfnamefont {R.}~\bibnamefont
  {Furfaro}}, \bibinfo {author} {\bibfnamefont {R.}~\bibnamefont {Linares}},\
  and\ \bibinfo {author} {\bibfnamefont {V.}~\bibnamefont {Reddy}},\ }\bibfield
   {title} {\bibinfo {title} {Shape identification of space objects via light
  curve inversion using deep learning models},\ }in\ \href
  {https://doi.org/https://amostech.com/TechnicalPapers/2019/Machine-Learning-for-SSA-Applications/Furfaro.pdf}
  {\emph {\bibinfo {booktitle} {AMOS Technologies Conference, Maui Economic
  Development Board, Kihei, Maui, HI}}}\ (\bibinfo {year} {2019})\BibitemShut
  {NoStop}%
\bibitem [{\citenamefont {Hearst}\ \emph {et~al.}(1998)\citenamefont {Hearst},
  \citenamefont {Dumais}, \citenamefont {Osuna}, \citenamefont {Platt},\ and\
  \citenamefont {Scholkopf}}]{hearst1998support}%
  \BibitemOpen
  \bibfield  {author} {\bibinfo {author} {\bibfnamefont {M.~A.}\ \bibnamefont
  {Hearst}}, \bibinfo {author} {\bibfnamefont {S.~T.}\ \bibnamefont {Dumais}},
  \bibinfo {author} {\bibfnamefont {E.}~\bibnamefont {Osuna}}, \bibinfo
  {author} {\bibfnamefont {J.}~\bibnamefont {Platt}},\ and\ \bibinfo {author}
  {\bibfnamefont {B.}~\bibnamefont {Scholkopf}},\ }\bibfield  {title} {\bibinfo
  {title} {Support vector machines},\ }\href
  {https://doi.org/10.1109/5254.708428} {\bibfield  {journal} {\bibinfo
  {journal} {IEEE Intelligent Systems and their applications}\ }\textbf
  {\bibinfo {volume} {13}},\ \bibinfo {pages} {18} (\bibinfo {year}
  {1998})}\BibitemShut {NoStop}%
\bibitem [{\citenamefont {Quinlan}(1987)}]{quinlan1987simplifying}%
  \BibitemOpen
  \bibfield  {author} {\bibinfo {author} {\bibfnamefont {J.~R.}\ \bibnamefont
  {Quinlan}},\ }\bibfield  {title} {\bibinfo {title} {Simplifying decision
  trees},\ }\href {https://doi.org/10.1016/S0020-7373(87)80053-6} {\bibfield
  {journal} {\bibinfo  {journal} {International journal of man-machine
  studies}\ }\textbf {\bibinfo {volume} {27}},\ \bibinfo {pages} {221}
  (\bibinfo {year} {1987})}\BibitemShut {NoStop}%
\bibitem [{\citenamefont {Le~Guennec}\ \emph {et~al.}(2016)\citenamefont
  {Le~Guennec}, \citenamefont {Malinowski},\ and\ \citenamefont
  {Tavenard}}]{LeGuennec2016}%
  \BibitemOpen
  \bibfield  {author} {\bibinfo {author} {\bibfnamefont {A.}~\bibnamefont
  {Le~Guennec}}, \bibinfo {author} {\bibfnamefont {S.}~\bibnamefont
  {Malinowski}},\ and\ \bibinfo {author} {\bibfnamefont {R.}~\bibnamefont
  {Tavenard}},\ }\bibfield  {title} {\bibinfo {title} {Data augmentation for
  time series classification using convolutional neural networks},\ }in\ \href
  {https://doi.org/https://shs.hal.science/halshs-01357973/} {\emph {\bibinfo
  {booktitle} {ECML/PKDD workshop on advanced analytics and learning on
  temporal data}}}\ (\bibinfo {year} {2016})\BibitemShut {NoStop}%
\bibitem [{\citenamefont {Wen}\ \emph {et~al.}(2020)\citenamefont {Wen},
  \citenamefont {Sun}, \citenamefont {Yang}, \citenamefont {Song},
  \citenamefont {Gao}, \citenamefont {Wang},\ and\ \citenamefont
  {Xu}}]{Wen2020}%
  \BibitemOpen
  \bibfield  {author} {\bibinfo {author} {\bibfnamefont {Q.}~\bibnamefont
  {Wen}}, \bibinfo {author} {\bibfnamefont {L.}~\bibnamefont {Sun}}, \bibinfo
  {author} {\bibfnamefont {F.}~\bibnamefont {Yang}}, \bibinfo {author}
  {\bibfnamefont {X.}~\bibnamefont {Song}}, \bibinfo {author} {\bibfnamefont
  {J.}~\bibnamefont {Gao}}, \bibinfo {author} {\bibfnamefont {X.}~\bibnamefont
  {Wang}},\ and\ \bibinfo {author} {\bibfnamefont {H.}~\bibnamefont {Xu}},\
  }\bibfield  {title} {\bibinfo {title} {Time series data augmentation for deep
  learning: A survey},\ }\bibfield  {journal} {\bibinfo  {journal}
  {arXiv:2002.12478}\ }\href {https://doi.org/https://arxiv.org/abs/2002.12478}
  {https://arxiv.org/abs/2002.12478} (\bibinfo {year} {2020})\BibitemShut
  {NoStop}%
\bibitem [{\citenamefont {Iwana}\ and\ \citenamefont
  {Uchida}(2021)}]{Iwana2020}%
  \BibitemOpen
  \bibfield  {author} {\bibinfo {author} {\bibfnamefont {B.~K.}\ \bibnamefont
  {Iwana}}\ and\ \bibinfo {author} {\bibfnamefont {S.}~\bibnamefont {Uchida}},\
  }\bibfield  {title} {\bibinfo {title} {An empirical survey of data
  augmentation for time series classification with neural networks},\ }\href
  {https://doi.org/10.1371/journal.pone.0254841} {\bibfield  {journal}
  {\bibinfo  {journal} {Plos one}\ }\textbf {\bibinfo {volume} {16}},\ \bibinfo
  {pages} {e0254841} (\bibinfo {year} {2021})}\BibitemShut {NoStop}%
\bibitem [{\citenamefont {Pearlman}\ \emph {et~al.}(2019)\citenamefont
  {Pearlman}, \citenamefont {Arnold}, \citenamefont {Davis}, \citenamefont
  {Barlier}, \citenamefont {Biancale}, \citenamefont {Vasiliev}, \citenamefont
  {Ciufolini}, \citenamefont {Paolozzi}, \citenamefont {Pavlis}, \citenamefont
  {So{{\'s}}nica} \emph {et~al.}}]{pearlman2019laser}%
  \BibitemOpen
  \bibfield  {author} {\bibinfo {author} {\bibfnamefont {M.}~\bibnamefont
  {Pearlman}}, \bibinfo {author} {\bibfnamefont {D.}~\bibnamefont {Arnold}},
  \bibinfo {author} {\bibfnamefont {M.}~\bibnamefont {Davis}}, \bibinfo
  {author} {\bibfnamefont {F.}~\bibnamefont {Barlier}}, \bibinfo {author}
  {\bibfnamefont {R.}~\bibnamefont {Biancale}}, \bibinfo {author}
  {\bibfnamefont {V.}~\bibnamefont {Vasiliev}}, \bibinfo {author}
  {\bibfnamefont {I.}~\bibnamefont {Ciufolini}}, \bibinfo {author}
  {\bibfnamefont {A.}~\bibnamefont {Paolozzi}}, \bibinfo {author}
  {\bibfnamefont {E.~C.}\ \bibnamefont {Pavlis}}, \bibinfo {author}
  {\bibfnamefont {K.}~\bibnamefont {So{{\'s}}nica}}, \emph {et~al.},\
  }\bibfield  {title} {\bibinfo {title} {Laser geodetic satellites: a
  high-accuracy scientific tool},\ }\href
  {https://doi.org/10.1007/s00190-019-01228-y} {\bibfield  {journal} {\bibinfo
  {journal} {Journal of Geodesy}\ }\textbf {\bibinfo {volume} {93}},\ \bibinfo
  {pages} {2181} (\bibinfo {year} {2019})}\BibitemShut {NoStop}%
\bibitem [{\citenamefont {Kucharski}\ \emph {et~al.}(2017)\citenamefont
  {Kucharski}, \citenamefont {Kirchner}, \citenamefont {Bennett}, \citenamefont
  {Lachut}, \citenamefont {So{\'s}nica}, \citenamefont {Koshkin}, \citenamefont
  {Shakun}, \citenamefont {Koidl}, \citenamefont {Steindorfer}, \citenamefont
  {Wang} \emph {et~al.}}]{kucharski2017}%
  \BibitemOpen
  \bibfield  {author} {\bibinfo {author} {\bibfnamefont {D.}~\bibnamefont
  {Kucharski}}, \bibinfo {author} {\bibfnamefont {G.}~\bibnamefont {Kirchner}},
  \bibinfo {author} {\bibfnamefont {J.}~\bibnamefont {Bennett}}, \bibinfo
  {author} {\bibfnamefont {M.}~\bibnamefont {Lachut}}, \bibinfo {author}
  {\bibfnamefont {K.}~\bibnamefont {So{\'s}nica}}, \bibinfo {author}
  {\bibfnamefont {N.}~\bibnamefont {Koshkin}}, \bibinfo {author} {\bibfnamefont
  {L.}~\bibnamefont {Shakun}}, \bibinfo {author} {\bibfnamefont
  {F.}~\bibnamefont {Koidl}}, \bibinfo {author} {\bibfnamefont
  {M.}~\bibnamefont {Steindorfer}}, \bibinfo {author} {\bibfnamefont
  {P.}~\bibnamefont {Wang}}, \emph {et~al.},\ }\bibfield  {title} {\bibinfo
  {title} {Photon pressure force on space debris topex/poseidon measured by
  satellite laser ranging},\ }\href {https://doi.org/10.1002/2017EA000329}
  {\bibfield  {journal} {\bibinfo  {journal} {Earth and Space Science}\
  }\textbf {\bibinfo {volume} {4}},\ \bibinfo {pages} {661} (\bibinfo {year}
  {2017})}\BibitemShut {NoStop}%
\bibitem [{\citenamefont {Kucharski}\ \emph {et~al.}(2014)\citenamefont
  {Kucharski}, \citenamefont {Kirchner}, \citenamefont {Koidl}, \citenamefont
  {Fan}, \citenamefont {Carman}, \citenamefont {Moore}, \citenamefont
  {Dmytrotsa}, \citenamefont {Ploner}, \citenamefont {Bianco}, \citenamefont
  {Medvedskij} \emph {et~al.}}]{kucharski2014}%
  \BibitemOpen
  \bibfield  {author} {\bibinfo {author} {\bibfnamefont {D.}~\bibnamefont
  {Kucharski}}, \bibinfo {author} {\bibfnamefont {G.}~\bibnamefont {Kirchner}},
  \bibinfo {author} {\bibfnamefont {F.}~\bibnamefont {Koidl}}, \bibinfo
  {author} {\bibfnamefont {C.}~\bibnamefont {Fan}}, \bibinfo {author}
  {\bibfnamefont {R.}~\bibnamefont {Carman}}, \bibinfo {author} {\bibfnamefont
  {C.}~\bibnamefont {Moore}}, \bibinfo {author} {\bibfnamefont
  {A.}~\bibnamefont {Dmytrotsa}}, \bibinfo {author} {\bibfnamefont
  {M.}~\bibnamefont {Ploner}}, \bibinfo {author} {\bibfnamefont
  {G.}~\bibnamefont {Bianco}}, \bibinfo {author} {\bibfnamefont
  {M.}~\bibnamefont {Medvedskij}}, \emph {et~al.},\ }\bibfield  {title}
  {\bibinfo {title} {Attitude and spin period of space debris envisat measured
  by satellite laser ranging},\ }\href
  {https://doi.org/10.1109/TGRS.2014.2316138} {\bibfield  {journal} {\bibinfo
  {journal} {IEEE Transactions on Geoscience and Remote Sensing}\ }\textbf
  {\bibinfo {volume} {52}},\ \bibinfo {pages} {7651} (\bibinfo {year}
  {2014})}\BibitemShut {NoStop}%
\bibitem [{\citenamefont {Kirchner}\ \emph {et~al.}(2017)\citenamefont
  {Kirchner}, \citenamefont {Steindorfer}, \citenamefont {Wang}, \citenamefont
  {Koidl}, \citenamefont {Kucharski}, \citenamefont {Silha}, \citenamefont
  {Schildknecht}, \citenamefont {Krag},\ and\ \citenamefont
  {Flohrer}}]{kirchner2017determination}%
  \BibitemOpen
  \bibfield  {author} {\bibinfo {author} {\bibfnamefont {G.}~\bibnamefont
  {Kirchner}}, \bibinfo {author} {\bibfnamefont {M.}~\bibnamefont
  {Steindorfer}}, \bibinfo {author} {\bibfnamefont {P.}~\bibnamefont {Wang}},
  \bibinfo {author} {\bibfnamefont {F.}~\bibnamefont {Koidl}}, \bibinfo
  {author} {\bibfnamefont {D.}~\bibnamefont {Kucharski}}, \bibinfo {author}
  {\bibfnamefont {J.}~\bibnamefont {Silha}}, \bibinfo {author} {\bibfnamefont
  {T.}~\bibnamefont {Schildknecht}}, \bibinfo {author} {\bibfnamefont
  {H.}~\bibnamefont {Krag}},\ and\ \bibinfo {author} {\bibfnamefont
  {T.}~\bibnamefont {Flohrer}},\ }\bibfield  {title} {\bibinfo {title}
  {Determination of attitude and attitude motion of space debris, using laser
  ranging and single-photon light curve data},\ }in\ \href@noop {} {\emph
  {\bibinfo {booktitle} {Proceedings of the 7th European Conference on Space
  Debris, Darmstadt, Germany}}}\ (\bibinfo {year} {2017})\ pp.\ \bibinfo
  {pages} {18--21}\BibitemShut {NoStop}%
\bibitem [{\citenamefont {Yu}\ and\ \citenamefont {Zhu}(2020)}]{yu2020hyper}%
  \BibitemOpen
  \bibfield  {author} {\bibinfo {author} {\bibfnamefont {T.}~\bibnamefont
  {Yu}}\ and\ \bibinfo {author} {\bibfnamefont {H.}~\bibnamefont {Zhu}},\
  }\bibfield  {title} {\bibinfo {title} {Hyper-parameter optimization: A review
  of algorithms and applications},\ }\bibfield  {journal} {\bibinfo  {journal}
  {arXiv:2003.05689}\ }\href {https://doi.org/10.48550/arXiv.2003.05689}
  {10.48550/arXiv.2003.05689} (\bibinfo {year} {2020})\BibitemShut {NoStop}%
\bibitem [{\citenamefont {Allworth}\ \emph {et~al.}(2020)\citenamefont
  {Allworth}, \citenamefont {Windrim}, \citenamefont {Wardman}, \citenamefont
  {Kucharski}, \citenamefont {Bennett},\ and\ \citenamefont
  {Bryson}}]{Allworth2020}%
  \BibitemOpen
  \bibfield  {author} {\bibinfo {author} {\bibfnamefont {J.}~\bibnamefont
  {Allworth}}, \bibinfo {author} {\bibfnamefont {L.}~\bibnamefont {Windrim}},
  \bibinfo {author} {\bibfnamefont {J.}~\bibnamefont {Wardman}}, \bibinfo
  {author} {\bibfnamefont {D.}~\bibnamefont {Kucharski}}, \bibinfo {author}
  {\bibfnamefont {J.}~\bibnamefont {Bennett}},\ and\ \bibinfo {author}
  {\bibfnamefont {M.}~\bibnamefont {Bryson}},\ }\bibfield  {title} {\bibinfo
  {title} {Development of a high fidelity simulator for generalised photometric
  based space object classification using machine learning},\ }\bibfield
  {journal} {\bibinfo  {journal} {arXiv:2004.12270}\ }\href
  {https://doi.org/https://arxiv.org/pdf/2004.12270}
  {https://arxiv.org/pdf/2004.12270} (\bibinfo {year} {2020})\BibitemShut
  {NoStop}%
\bibitem [{\citenamefont {Cunningham}\ and\ \citenamefont
  {Delany}(2021)}]{cunningham2021k}%
  \BibitemOpen
  \bibfield  {author} {\bibinfo {author} {\bibfnamefont {P.}~\bibnamefont
  {Cunningham}}\ and\ \bibinfo {author} {\bibfnamefont {S.~J.}\ \bibnamefont
  {Delany}},\ }\bibfield  {title} {\bibinfo {title} {K-nearest neighbour
  classifiers-a tutorial},\ }\href {https://doi.org/10.1145/3459665} {\bibfield
   {journal} {\bibinfo  {journal} {ACM computing surveys (CSUR)}\ }\textbf
  {\bibinfo {volume} {54}},\ \bibinfo {pages} {1} (\bibinfo {year}
  {2021})}\BibitemShut {NoStop}%
\bibitem [{\citenamefont {Pal}(2005)}]{pal2005random}%
  \BibitemOpen
  \bibfield  {author} {\bibinfo {author} {\bibfnamefont {M.}~\bibnamefont
  {Pal}},\ }\bibfield  {title} {\bibinfo {title} {Random forest classifier for
  remote sensing classification},\ }\href
  {https://doi.org/10.1080/01431160412331269698} {\bibfield  {journal}
  {\bibinfo  {journal} {International journal of remote sensing}\ }\textbf
  {\bibinfo {volume} {26}},\ \bibinfo {pages} {217} (\bibinfo {year}
  {2005})}\BibitemShut {NoStop}%
\bibitem [{\citenamefont {Chen}\ and\ \citenamefont
  {Guestrin}(2016)}]{chen2016xgboost}%
  \BibitemOpen
  \bibfield  {author} {\bibinfo {author} {\bibfnamefont {T.}~\bibnamefont
  {Chen}}\ and\ \bibinfo {author} {\bibfnamefont {C.}~\bibnamefont
  {Guestrin}},\ }\bibfield  {title} {\bibinfo {title} {Xgboost: A scalable tree
  boosting system},\ }in\ \href {https://doi.org/10.1145/2939672.2939785}
  {\emph {\bibinfo {booktitle} {Proceedings of the 22nd acm sigkdd
  international conference on knowledge discovery and data mining}}}\ (\bibinfo
  {year} {2016})\ pp.\ \bibinfo {pages} {785--794}\BibitemShut {NoStop}%
\bibitem [{\citenamefont {Abanda}\ \emph {et~al.}(2019)\citenamefont {Abanda},
  \citenamefont {Mori},\ and\ \citenamefont {Lozano}}]{Abanda2019}%
  \BibitemOpen
  \bibfield  {author} {\bibinfo {author} {\bibfnamefont {A.}~\bibnamefont
  {Abanda}}, \bibinfo {author} {\bibfnamefont {U.}~\bibnamefont {Mori}},\ and\
  \bibinfo {author} {\bibfnamefont {J.~A.}\ \bibnamefont {Lozano}},\ }\bibfield
   {title} {\bibinfo {title} {A review on distance based time series
  classification},\ }\href {https://doi.org/10.1007/s10618-018-0596-4}
  {\bibfield  {journal} {\bibinfo  {journal} {Data Mining and Knowledge
  Discovery}\ }\textbf {\bibinfo {volume} {33}},\ \bibinfo {pages} {378}
  (\bibinfo {year} {2019})}\BibitemShut {NoStop}%
\bibitem [{\citenamefont {Erickson}\ and\ \citenamefont
  {Kitamura}(2021)}]{erickson2021magician}%
  \BibitemOpen
  \bibfield  {author} {\bibinfo {author} {\bibfnamefont {B.~J.}\ \bibnamefont
  {Erickson}}\ and\ \bibinfo {author} {\bibfnamefont {F.}~\bibnamefont
  {Kitamura}},\ }\href {https://doi.org/10.1148/ryai.2021200126} {\bibinfo
  {title} {Magician’s corner: 9. performance metrics for machine learning
  models}} (\bibinfo {year} {2021})\BibitemShut {NoStop}%
\bibitem [{\citenamefont {Grandini}\ \emph {et~al.}(2020)\citenamefont
  {Grandini}, \citenamefont {Bagli},\ and\ \citenamefont
  {Visani}}]{grandini2020metrics}%
  \BibitemOpen
  \bibfield  {author} {\bibinfo {author} {\bibfnamefont {M.}~\bibnamefont
  {Grandini}}, \bibinfo {author} {\bibfnamefont {E.}~\bibnamefont {Bagli}},\
  and\ \bibinfo {author} {\bibfnamefont {G.}~\bibnamefont {Visani}},\
  }\bibfield  {title} {\bibinfo {title} {Metrics for multi-class
  classification: an overview},\ }\bibfield  {journal} {\bibinfo  {journal}
  {arXiv:2008.05756}\ }\href {https://doi.org/10.48550/arXiv.2008.05756}
  {10.48550/arXiv.2008.05756} (\bibinfo {year} {2020})\BibitemShut {NoStop}%
\bibitem [{\citenamefont {Wilkinson}\ \emph {et~al.}(2019)\citenamefont
  {Wilkinson}, \citenamefont {Schreiber}, \citenamefont {Proch{{\'a}}zka},
  \citenamefont {Moore}, \citenamefont {Degnan}, \citenamefont {Kirchner},
  \citenamefont {Zhongping}, \citenamefont {Dunn}, \citenamefont
  {Shargorodskiy}, \citenamefont {Sadovnikov} \emph {et~al.}}]{Wilkinson2019}%
  \BibitemOpen
  \bibfield  {author} {\bibinfo {author} {\bibfnamefont {M.}~\bibnamefont
  {Wilkinson}}, \bibinfo {author} {\bibfnamefont {U.}~\bibnamefont
  {Schreiber}}, \bibinfo {author} {\bibfnamefont {I.}~\bibnamefont
  {Proch{{\'a}}zka}}, \bibinfo {author} {\bibfnamefont {C.}~\bibnamefont
  {Moore}}, \bibinfo {author} {\bibfnamefont {J.}~\bibnamefont {Degnan}},
  \bibinfo {author} {\bibfnamefont {G.}~\bibnamefont {Kirchner}}, \bibinfo
  {author} {\bibfnamefont {Z.}~\bibnamefont {Zhongping}}, \bibinfo {author}
  {\bibfnamefont {P.}~\bibnamefont {Dunn}}, \bibinfo {author} {\bibfnamefont
  {V.}~\bibnamefont {Shargorodskiy}}, \bibinfo {author} {\bibfnamefont
  {M.}~\bibnamefont {Sadovnikov}}, \emph {et~al.},\ }\bibfield  {title}
  {\bibinfo {title} {The next generation of satellite laser ranging systems},\
  }\href {https://doi.org/10.1007/s00190-018-1196-1} {\bibfield  {journal}
  {\bibinfo  {journal} {Journal of Geodesy}\ }\textbf {\bibinfo {volume}
  {93}},\ \bibinfo {pages} {2227} (\bibinfo {year} {2019})}\BibitemShut
  {NoStop}%
\bibitem [{\citenamefont {Steindorfer}\ \emph {et~al.}(2020)\citenamefont
  {Steindorfer}, \citenamefont {Kirchner}, \citenamefont {Koidl}, \citenamefont
  {Wang}, \citenamefont {Jilete},\ and\ \citenamefont
  {Flohrer}}]{Steindorfer2020}%
  \BibitemOpen
  \bibfield  {author} {\bibinfo {author} {\bibfnamefont {M.~A.}\ \bibnamefont
  {Steindorfer}}, \bibinfo {author} {\bibfnamefont {G.}~\bibnamefont
  {Kirchner}}, \bibinfo {author} {\bibfnamefont {F.}~\bibnamefont {Koidl}},
  \bibinfo {author} {\bibfnamefont {P.}~\bibnamefont {Wang}}, \bibinfo {author}
  {\bibfnamefont {B.}~\bibnamefont {Jilete}},\ and\ \bibinfo {author}
  {\bibfnamefont {T.}~\bibnamefont {Flohrer}},\ }\bibfield  {title} {\bibinfo
  {title} {Daylight space debris laser ranging},\ }\href
  {https://doi.org/10.1038/s41467-020-17332-z} {\bibfield  {journal} {\bibinfo
  {journal} {Nature communications}\ }\textbf {\bibinfo {volume} {11}},\
  \bibinfo {pages} {3735} (\bibinfo {year} {2020})}\BibitemShut {NoStop}%
\bibitem [{\citenamefont {Trimberger}(2012)}]{trimberger2012field}%
  \BibitemOpen
  \bibfield  {author} {\bibinfo {author} {\bibfnamefont {S.~M.}\ \bibnamefont
  {Trimberger}},\ }\href@noop {} {\emph {\bibinfo {title} {Field-programmable
  gate array technology}}}\ (\bibinfo  {publisher} {Springer Science \&
  Business Media},\ \bibinfo {year} {2012})\BibitemShut {NoStop}%
\bibitem [{\citenamefont {Steindorfer}\ \emph {et~al.}(2015)\citenamefont
  {Steindorfer}, \citenamefont {Kirchner}, \citenamefont {Koidl},\ and\
  \citenamefont {Wang}}]{Steindorfer2015}%
  \BibitemOpen
  \bibfield  {author} {\bibinfo {author} {\bibfnamefont {M.}~\bibnamefont
  {Steindorfer}}, \bibinfo {author} {\bibfnamefont {G.}~\bibnamefont
  {Kirchner}}, \bibinfo {author} {\bibfnamefont {F.}~\bibnamefont {Koidl}},\
  and\ \bibinfo {author} {\bibfnamefont {P.}~\bibnamefont {Wang}},\ }\bibfield
  {title} {\bibinfo {title} {Light curve measurements with single photon
  counters at graz slr},\ }in\ \href@noop {} {\emph {\bibinfo {booktitle} {2015
  ILRS Technical Workshop}}}\ (\bibinfo {year} {2015})\ pp.\ \bibinfo {pages}
  {1--7}\BibitemShut {NoStop}%
\bibitem [{\citenamefont {Bostrom}\ and\ \citenamefont
  {Bagnall}(2015)}]{bostrom2015}%
  \BibitemOpen
  \bibfield  {author} {\bibinfo {author} {\bibfnamefont {A.}~\bibnamefont
  {Bostrom}}\ and\ \bibinfo {author} {\bibfnamefont {A.}~\bibnamefont
  {Bagnall}},\ }\bibfield  {title} {\bibinfo {title} {Binary shapelet transform
  for multiclass time series classification},\ }in\ \href@noop {} {\emph
  {\bibinfo {booktitle} {Big Data Analytics and Knowledge Discovery: 17th
  International Conference, DaWaK 2015, Valencia, Spain, September 1-4, 2015,
  Proceedings 17}}}\ (\bibinfo {organization} {Springer},\ \bibinfo {year}
  {2015})\ pp.\ \bibinfo {pages} {257--269}\BibitemShut {NoStop}%
\bibitem [{\citenamefont {Christ}\ \emph {et~al.}(2016)\citenamefont {Christ},
  \citenamefont {Kempa-Liehr},\ and\ \citenamefont {Feindt}}]{Christ2016}%
  \BibitemOpen
  \bibfield  {author} {\bibinfo {author} {\bibfnamefont {M.}~\bibnamefont
  {Christ}}, \bibinfo {author} {\bibfnamefont {A.~W.}\ \bibnamefont
  {Kempa-Liehr}},\ and\ \bibinfo {author} {\bibfnamefont {M.}~\bibnamefont
  {Feindt}},\ }\bibfield  {title} {\bibinfo {title} {Distributed and parallel
  time series feature extraction for industrial big data applications},\
  }\bibfield  {journal} {\bibinfo  {journal} {arXiv preprint arXiv:1610.07717}\
  }\href {https://doi.org/https://arxiv.org/abs/1610.07717}
  {https://arxiv.org/abs/1610.07717} (\bibinfo {year} {2016})\BibitemShut
  {NoStop}%
\bibitem [{\citenamefont {Wang}\ \emph {et~al.}(2013)\citenamefont {Wang},
  \citenamefont {Mueen}, \citenamefont {Ding}, \citenamefont {Trajcevski},
  \citenamefont {Scheuermann},\ and\ \citenamefont {Keogh}}]{Wang2013}%
  \BibitemOpen
  \bibfield  {author} {\bibinfo {author} {\bibfnamefont {X.}~\bibnamefont
  {Wang}}, \bibinfo {author} {\bibfnamefont {A.}~\bibnamefont {Mueen}},
  \bibinfo {author} {\bibfnamefont {H.}~\bibnamefont {Ding}}, \bibinfo {author}
  {\bibfnamefont {G.}~\bibnamefont {Trajcevski}}, \bibinfo {author}
  {\bibfnamefont {P.}~\bibnamefont {Scheuermann}},\ and\ \bibinfo {author}
  {\bibfnamefont {E.}~\bibnamefont {Keogh}},\ }\bibfield  {title} {\bibinfo
  {title} {Experimental comparison of representation methods and distance
  measures for time series data},\ }\href
  {https://doi.org/10.1007/s10618-012-0250-5} {\bibfield  {journal} {\bibinfo
  {journal} {Data Mining and Knowledge Discovery}\ }\textbf {\bibinfo {volume}
  {26}},\ \bibinfo {pages} {275} (\bibinfo {year} {2013})}\BibitemShut
  {NoStop}%
\bibitem [{\citenamefont {Lines}\ and\ \citenamefont
  {Bagnall}(2015)}]{Lines2015}%
  \BibitemOpen
  \bibfield  {author} {\bibinfo {author} {\bibfnamefont {J.}~\bibnamefont
  {Lines}}\ and\ \bibinfo {author} {\bibfnamefont {A.}~\bibnamefont
  {Bagnall}},\ }\bibfield  {title} {\bibinfo {title} {Time series
  classification with ensembles of elastic distance measures},\ }\href
  {https://doi.org/10.1007/s10618-014-0361-2} {\bibfield  {journal} {\bibinfo
  {journal} {Data Mining and Knowledge Discovery}\ }\textbf {\bibinfo {volume}
  {29}},\ \bibinfo {pages} {565} (\bibinfo {year} {2015})}\BibitemShut
  {NoStop}%
\bibitem [{\citenamefont {Sch{\"a}fer}(2016)}]{Schafer2016}%
  \BibitemOpen
  \bibfield  {author} {\bibinfo {author} {\bibfnamefont {P.}~\bibnamefont
  {Sch{\"a}fer}},\ }\bibfield  {title} {\bibinfo {title} {Scalable time series
  classification},\ }\href {https://doi.org/10.1007/s10618-015-0441-y}
  {\bibfield  {journal} {\bibinfo  {journal} {Data Mining and Knowledge
  Discovery}\ }\textbf {\bibinfo {volume} {30}},\ \bibinfo {pages} {1273}
  (\bibinfo {year} {2016})}\BibitemShut {NoStop}%
\bibitem [{\citenamefont {Greenacre}\ \emph {et~al.}(2022)\citenamefont
  {Greenacre}, \citenamefont {Groenen}, \citenamefont {Hastie}, \citenamefont
  {d’Enza}, \citenamefont {Markos},\ and\ \citenamefont
  {Tuzhilina}}]{greenacre2022principal}%
  \BibitemOpen
  \bibfield  {author} {\bibinfo {author} {\bibfnamefont {M.}~\bibnamefont
  {Greenacre}}, \bibinfo {author} {\bibfnamefont {P.~J.}\ \bibnamefont
  {Groenen}}, \bibinfo {author} {\bibfnamefont {T.}~\bibnamefont {Hastie}},
  \bibinfo {author} {\bibfnamefont {A.~I.}\ \bibnamefont {d’Enza}}, \bibinfo
  {author} {\bibfnamefont {A.}~\bibnamefont {Markos}},\ and\ \bibinfo {author}
  {\bibfnamefont {E.}~\bibnamefont {Tuzhilina}},\ }\bibfield  {title} {\bibinfo
  {title} {Principal component analysis},\ }\href
  {https://doi.org/https://www.nature.com/articles/s43586-022-00184-w}
  {\bibfield  {journal} {\bibinfo  {journal} {Nature Reviews Methods Primers}\
  }\textbf {\bibinfo {volume} {2}},\ \bibinfo {pages} {100} (\bibinfo {year}
  {2022})}\BibitemShut {NoStop}%
\bibitem [{\citenamefont {Pedregosa}\ \emph {et~al.}(2011)\citenamefont
  {Pedregosa}, \citenamefont {Varoquaux}, \citenamefont {Gramfort},
  \citenamefont {Michel}, \citenamefont {Thirion}, \citenamefont {Grisel},
  \citenamefont {Blondel}, \citenamefont {Prettenhofer}, \citenamefont {Weiss},
  \citenamefont {Dubourg} \emph {et~al.}}]{pedregosa2011scikit}%
  \BibitemOpen
  \bibfield  {author} {\bibinfo {author} {\bibfnamefont {F.}~\bibnamefont
  {Pedregosa}}, \bibinfo {author} {\bibfnamefont {G.}~\bibnamefont
  {Varoquaux}}, \bibinfo {author} {\bibfnamefont {A.}~\bibnamefont {Gramfort}},
  \bibinfo {author} {\bibfnamefont {V.}~\bibnamefont {Michel}}, \bibinfo
  {author} {\bibfnamefont {B.}~\bibnamefont {Thirion}}, \bibinfo {author}
  {\bibfnamefont {O.}~\bibnamefont {Grisel}}, \bibinfo {author} {\bibfnamefont
  {M.}~\bibnamefont {Blondel}}, \bibinfo {author} {\bibfnamefont
  {P.}~\bibnamefont {Prettenhofer}}, \bibinfo {author} {\bibfnamefont
  {R.}~\bibnamefont {Weiss}}, \bibinfo {author} {\bibfnamefont
  {V.}~\bibnamefont {Dubourg}}, \emph {et~al.},\ }\bibfield  {title} {\bibinfo
  {title} {Scikit-learn: Machine learning in python},\ }\href
  {https://doi.org/https://www.jmlr.org/papers/volume12/pedregosa11a/pedregosa11a.pdf?ref=https:/}
  {\bibfield  {journal} {\bibinfo  {journal} {the Journal of machine Learning
  research}\ }\textbf {\bibinfo {volume} {12}},\ \bibinfo {pages} {2825}
  (\bibinfo {year} {2011})}\BibitemShut {NoStop}%
\bibitem [{\citenamefont {Chen}\ \emph {et~al.}(2019)\citenamefont {Chen},
  \citenamefont {He}, \citenamefont {Benesty},\ and\ \citenamefont
  {Khotilovich}}]{chen2019package}%
  \BibitemOpen
  \bibfield  {author} {\bibinfo {author} {\bibfnamefont {T.}~\bibnamefont
  {Chen}}, \bibinfo {author} {\bibfnamefont {T.}~\bibnamefont {He}}, \bibinfo
  {author} {\bibfnamefont {M.}~\bibnamefont {Benesty}},\ and\ \bibinfo {author}
  {\bibfnamefont {V.}~\bibnamefont {Khotilovich}},\ }\bibfield  {title}
  {\bibinfo {title} {Package ‘xgboost’},\ }\href
  {https://doi.org/http://r.meteo.uni.wroc.pl/web/packages/xgboost/xgboost.pdf}
  {\bibfield  {journal} {\bibinfo  {journal} {R version}\ }\textbf {\bibinfo
  {volume} {90}},\ \bibinfo {pages} {40} (\bibinfo {year} {2019})}\BibitemShut
  {NoStop}%
\bibitem [{\citenamefont {Abadi}\ \emph {et~al.}(2015)\citenamefont {Abadi},
  \citenamefont {Agarwal}, \citenamefont {Barham}, \citenamefont {Brevdo},
  \citenamefont {Chen}, \citenamefont {Citro}, \citenamefont {Corrado},
  \citenamefont {Davis}, \citenamefont {Dean}, \citenamefont {Devin} \emph
  {et~al.}}]{tensorflow2015-whitepaper}%
  \BibitemOpen
  \bibfield  {author} {\bibinfo {author} {\bibfnamefont {M.}~\bibnamefont
  {Abadi}}, \bibinfo {author} {\bibfnamefont {A.}~\bibnamefont {Agarwal}},
  \bibinfo {author} {\bibfnamefont {P.}~\bibnamefont {Barham}}, \bibinfo
  {author} {\bibfnamefont {E.}~\bibnamefont {Brevdo}}, \bibinfo {author}
  {\bibfnamefont {Z.}~\bibnamefont {Chen}}, \bibinfo {author} {\bibfnamefont
  {C.}~\bibnamefont {Citro}}, \bibinfo {author} {\bibfnamefont {G.~S.}\
  \bibnamefont {Corrado}}, \bibinfo {author} {\bibfnamefont {A.}~\bibnamefont
  {Davis}}, \bibinfo {author} {\bibfnamefont {J.}~\bibnamefont {Dean}},
  \bibinfo {author} {\bibfnamefont {M.}~\bibnamefont {Devin}}, \emph {et~al.},\
  }\href {https://doi.org/10.48550/arXiv.1603.04467} {\bibinfo {title}
  {Tensorflow: Large-scale machine learning on heterogeneous systems}}
  (\bibinfo {year} {2015})\BibitemShut {NoStop}%
\bibitem [{\citenamefont {Qashoa}\ and\ \citenamefont
  {Lee}(2023)}]{Qashoa2023}%
  \BibitemOpen
  \bibfield  {author} {\bibinfo {author} {\bibfnamefont {R.}~\bibnamefont
  {Qashoa}}\ and\ \bibinfo {author} {\bibfnamefont {R.}~\bibnamefont {Lee}},\
  }\bibfield  {title} {\bibinfo {title} {Classification of low earth orbit
  (leo) resident space objects’(rso) light curves using a support vector
  machine (svm) and long short-term memory (lstm)},\ }\href
  {https://doi.org/10.3390/s23146539} {\bibfield  {journal} {\bibinfo
  {journal} {Sensors}\ }\textbf {\bibinfo {volume} {23}},\ \bibinfo {pages}
  {6539} (\bibinfo {year} {2023})}\BibitemShut {NoStop}%
\end{thebibliography}%
\newpage{}
\section*{Appendix} 
\label{sec:appendix}
\begin{figure}[h!]
\centering
\includegraphics[width=0.5\textwidth]{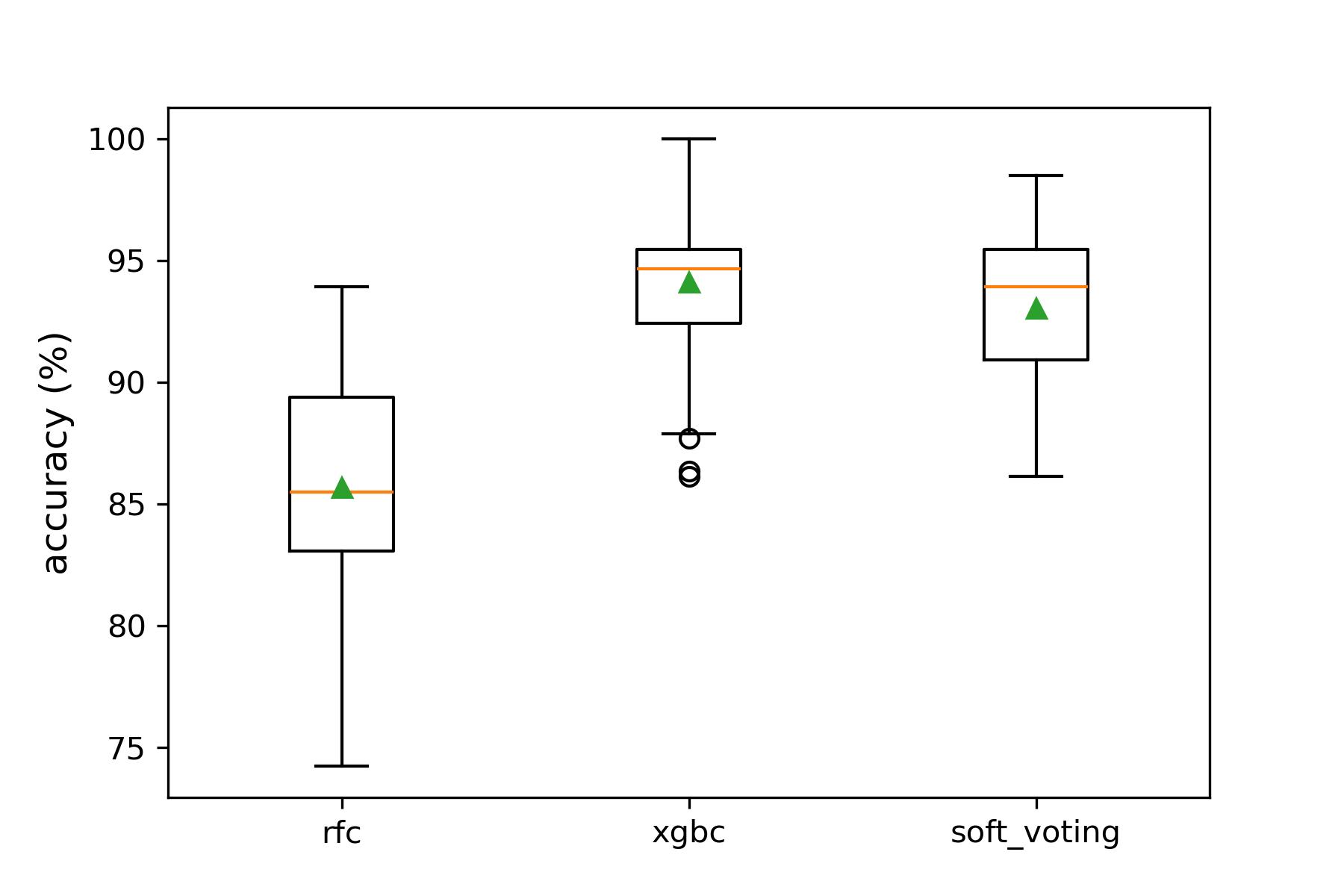}
\caption{Ensemble of random forest (rfc) and xgboost (xgc) classifiers for the classification of \textit{Families} task with soft-voting scheme. }
\label{fig:ensemble_soft}
\end{figure}

The results show that the accuracy of random forest and xgboost classifiers is better than that of other classifier methods. Hence, we have further explored the ensembling of these two classifiers to perform classification on the same data. An ensemble of classifiers could be possible in several ways (e.g., max voting, stacking, blending, bagging, boosting). A majority voting ensemble combines the prediction from random forest and xgboost classifiers. We can determine the majority vote for classification problems as soft and hard voting. Soft voting involves predicting the class with the most significant probability. On the other hand, hard voting predicts the class with the most number of votes. 
Figures \ref{fig:ensemble_soft} and \ref{fig:ensemble_hard} show the ensemble results for the ``Families" category with soft and hard voting. For soft voting, the testing accuracy with individual classifiers (10-fold) is $85.7 \%$ and $94.1 \%$, respectively, whereas the accuracy of the ensembled classifier is $93\%$. Similarly, for hard voting, the testing accuracy for the random forest is $91\%$, and for xgboost is $94.1\%$. The ensemble classifier shows an accuracy of $94.2\%$. Hyperparameter optimization configures the parameters for the random forest and xgboost classifiers. We performed separate grid and random searches for the hyperparameter optimization for these two classifiers. 
The search space for random forest is designed based on three specific parameters: number of estimators, maximum depth, and maximum leaf nodes. We chose $[195, 220]$, $[3, 9]$, and $[3, 9]$ as the prior boundaries for these parameters. Both grid and random searches provide the best values for the number of estimators, maximum depth, and maximum leaf nodes as $220$, $9$, and $9$. However, a broader range of prior boundaries could provide another optimal value of the parameters. Notably, the computational cost of performing a grid search with large training data is expensive. For example, for this exercise, a grid-based search takes $ \sim 17$ minutes with 526 time series data with $342$ features, whereas a random search takes $\sim 1$ minute. Hence, random search would make exploring the optimal hyperparameters with more extensive boundaries more feasible. However, the random search would not guarantee the best set of hyperparameters, and in every run, it would give different solutions. A grid search is more robust than a random search and provides the best combination of hyperparameters with higher probabilities. This means there is always a trade-off between computational cost and the best hyperparameter finding in a grid search with large training data. This work does not suffer from this issue as we have a moderate amount of training data.  
Similarly, for the xgboost classifier, we performed a grid and random search with parameters such as the number of estimators, learning rate, and maximum depth. We chose a boundary between $0.1$ and $0.4$ for the learning rate, whereas the parameter boundaries are the same as the random forest classifier. This exercise shows that ensemble learning would improve the classification accuracy of our data, and we will explore this possibility in detail in our follow-up work. 
All these computational cost related experiments were performed on $\texttt{13th Gen Intel(R) Core(TM) i9-13900}$ processors.

\begin{figure}[h!]
\centering
\includegraphics[width=0.5\textwidth]{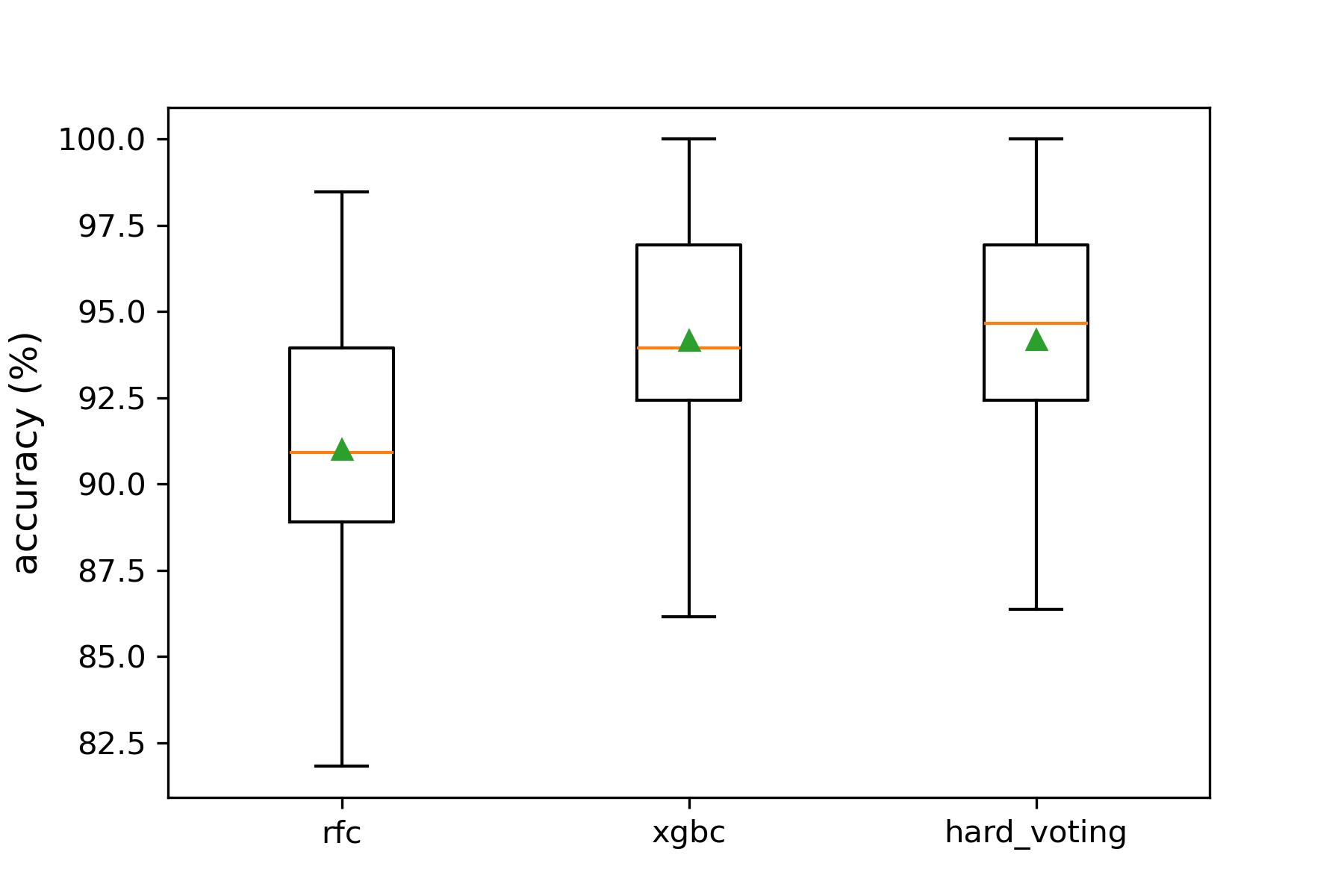}
\caption{Ensemble of random forest (rfc) and xgboost (xgc) classifiers for the classification of \textit{Families} task with hard-voting scheme.}
\label{fig:ensemble_hard}
\end{figure}

\end{document}